\documentclass{ipi}
\usepackage{amsmath}
\usepackage{amsfonts}
\usepackage{amssymb}
  \usepackage{paralist}
  \usepackage{graphics} 
  \usepackage{epsfig} 
\usepackage{graphicx}  \usepackage{epstopdf}
 \usepackage[colorlinks=true]{hyperref}
\hypersetup{urlcolor=blue, citecolor=red}

\def\currentvolume{X}
 \def\currentissue{X}
  \def\currentyear{20xx}
   
    \def\ppages{X--XX}
     \def\DOI{10.3934/ipi.xx.xx.xx}

\newtheorem{theorem}{Theorem}[section]

\newtheorem{lemma}[theorem]{Lemma}
\newtheorem{proposition}{Proposition}

\theoremstyle{definition}

\DeclareMathAlphabet{\itbf}{OML}{cmm}{b}{it}

\def\br{{{\itbf r}}}

\def\by{{{\itbf y}}}
\def\bx{{{\itbf x}}}
\def\bz{{{\itbf z}}}

\def\bz{{{\itbf z}}}

\def\EE{{\mathbb{E}}}

\def\RR{{\mathbb{R}}}

\title{Ambient noise correlation-based imaging with moving sensors}

\author[Mathias Fink and Josselin Garnier]{}


\email{mathias.fink@espci.fr}
\email{josselin.garnier@polytechnique.edu}

\thanks{This work was supported by LABEX WIFI (Laboratory of Excellence ANR-10-LABX-24) within the French Program Investments for the Future under reference ANR-10-IDEX-0001-02 PSL*, 
and by ANR project SURMITO}

\begin{document}
\maketitle

\centerline{\scshape Mathias Fink}
\medskip
{\footnotesize 
 \centerline{Institut Langevin,  ESPCI and CNRS, PSL Research University}
 \centerline{ 1 rue Jussieu, 75005 Paris, France}
}

\medskip

\centerline{\scshape Josselin Garnier\footnote[1]{Corresponding author}}
\medskip
{\footnotesize 
 \centerline{Centre de Math\'ematiques Appliqu\'ees,
Ecole Polytechnique}
 \centerline{91128 Palaiseau Cedex,
France}
}

\bigskip

 \centerline{(Communicated by the associate editor name)}

\begin{abstract}
Waves can be used to probe and image an unknown medium.
Passive imaging uses ambient noise sources to illuminate the medium.
This paper considers passive imaging with moving sensors. The motivation is to generate large synthetic apertures,
which should result in enhanced resolution. 
However Doppler effects and lack of reciprocity
significantly affect the imaging process.
This paper discusses the consequences in terms of resolution 
and it shows how to design appropriate imaging functions depending on the sensor trajectory and velocity.  
\end{abstract}

\section{Introduction}
It is now well-known that the Green's function of the wave equation can be estimated 
from the cross correlation of the signals emitted by ambient noise sources and recorded by passive sensors
 \cite{badon15,bardos08,campillo03,colin09,davy,garnier05,schuster,wap04,weaver}.
In a homogeneous medium and when the source of the waves is a space-time stationary random field
that is also delta-correlated in space and time, it has been shown
\cite{snieder04,roux05}
that the derivative of the cross correlation of the signals recorded by two sensors is proportional to
the symmetrized Green's function between the sensors.
In an inhomogeneous medium and when the sources completely surround the region of the sensors
it can be shown using the Helmholtz-Kirchhoff identity that
there is a relation between the cross correlation of the recorded signals and the Green's function
\cite{wap10,garnier09}.
This is true even with spatially localized noise source distributions provided the waves 
propagate within an ergodic cavity \cite{bardos08}.
More generally, in an inhomogeneous medium the cross correlation as a function of the lag time
can have a distinguishable peak at  plus or minus the inter-sensor travel time, 
provided the ambient noise sources are well distributed around the sensors. 
The inter-sensor travel times obtained from peaks of cross correlations
can then be used tomographically for background velocity
estimation \cite{brenguier08,gouedard08,shapiro05,dehoop06}.
Additional peaks due to reflectors can be exploited so that reflectors 
can be imaged by migration of the cross correlation matrix 
of the signals emitted by ambient noise sources and recorded by a passive receiver array \cite{garnier09,garnier10,gouedard08}.
In this paper we extend these results to situations in which the receivers are moving.
The use of moving receivers is motivated by the general result that resolution is better when
the receiver array is large.
Since large physical arrays are difficult to implement, a natural 
idea is to implement moving sensors to generate large synthetic apertures.
So far very few results are available in this direction. Only Sabra mentions that 
Doppler effects should not affect Green's
function estimation from ambient noise cross-correlations
in underwater acoustics, when the sensors are moving with a velocity of a few meters
per second (which is very small compared to the sound speed that is approximately 1500 meters per second) \cite{sabra10}.
In \cite{garfink} a different but related problem is addressed: the  analysis of time-reversal experiments involving a moving point source that emits a pulse.
It is shown that Doppler effects and lack of source-receiver reciprocity significantly affect the time-reversal refocusing when the velocity of the source becomes comparable as the speed of propagation and refocusing can be enhanced by these effects.
Indeed the source-receiver reciprocity property means that the recorded signal is not modified
if we interchange the source and the receiver, and this comes from the symmetry of the Green's function.
However this reciprocity is broken when the source moves. It is also broken when the receiver moves. 
As we will see Doppler effects and  lack of reciprocity  also significantly affect correlation-based imaging when the sensor velocity
is comparable to the wave speed, but here resolution is reduced.

We will consider the following situation in the two-dimensional set-up in Sections \ref{sec:gre}-\ref{sec:ref}:
Noise sources are at the surface of a large ball and emit stationary random signals. 
A receiver is moving along a circular trajectory and records the field. 
The medium may be complex within the circular trajectory of the receiver (see Figure \ref{fig1a}). 
It is shown that the autocorrelation function of the recorded signal is related to the matrix of Green's function between pairs of points along the trajectory,
more exactly to a diagonal band of this matrix whose thickness is determined by the velocity of the receiver.
As an application we consider the case where a point-like reflector is present within the circular trajectory of the receiver (see Figure \ref{fig1})
and we show how to use the autocorrelation function  of the recorded signal to localize the reflector by migration.
A first naive migration function is proposed. Its analysis reveals that it has a strong bias and that 
a modification is needed when the velocity of the moving receiver 
is not negligible compared to the speed of propagation.
By applying this modification one gets an imaging function whose bias is negligible (i.e. smaller than the wavelength) 
but whose resolution (of the order of the wavelength) is reduced when the velocity of the receiver increases. 
A variant of this situation in which the  receiver is moving along a linear trajectory is addressed in Section \ref{sec:ref2} (see Figure \ref{fig3}). 
The analysis and conclusions are analogous to the case of a circular trajectory.

By the same strategy it is possible to study other types of situations related to passive Green's function estimation with moving objects, 
when the sources themselves are moving. 
In Section \ref{sec:1} we consider the case in which the ambient noise is emitted by a point-like source that
moves along a circular trajectory and that emits a stationary random signal.
Two observation points within the circle record the field (see Figure \ref{fig0}). 
The recorded signals are cross correlated.
It is shown that the cross correlation of the recorded signals 
is close to the Green's function between the two observation points,
although a correction appears when the velocity of the moving source is large and the noise bandwidth is limited.
However this correction vanishes for time lags approximately equal to the travel time between the 
sensors, which means that travel time estimation can be carried out 
with a bias smaller than the resolution and 
with the same resolution as if we were measuring the impulse response at one sensor
when the other one emits a pulse with the same spectrum as the power spectral density of the noise source.

\section{Passive Green's function estimation from a receiver moving on a circular trajectory}
\label{sec:gre}%
The goal of this section is to show that the autocorrelation function of the signal emitted by ambient noise sources and
recorded by a unique receiver moving on a circular trajectory is related to the Green's functions between pairs of points along the circular trajectory. This will be used in the next section to localize a reflector embedded in the medium.\\

\begin{figure}
\begin{center}
\begin{tabular}{c}
 \psfig{file=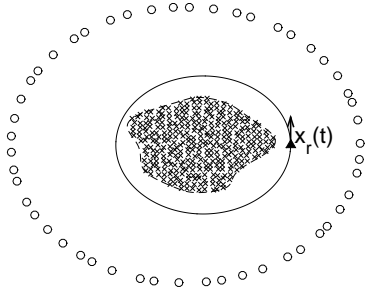,width=6.0cm}
\end{tabular}
\end{center}
\caption{Experimental set-up for passive Green's function estimation in Section \ref{sec:gre}. The circles are noise sources (at the surface $\partial B$), 
the triangle is a receiver at ${\bx}_{\rm r}(t)$ on a circular trajectory (with radius $R_0$), and the shaded area is a complex medium.
\label{fig1a}}
\end{figure}

{\bf Experimental set-up.}
We consider a  receiver moving on a circular trajectory at constant velocity.
Its position is~\cite{pied}:
\begin{equation}
{\bx}_{\rm r}(t) = (R_0 \cos(vt), R_0 \sin(vt)), 
\end{equation}
where $R_0>0$ is the radius of its circular trajectory and $v$ is its angular velocity (its linear velocity is $v_0=vR_0$).
Ambient noise sources located at the surface of a large ball $B$ emit stationary random signals
(the ball $B$ does not need to be centered at ${\bf 0}$, but it needs to enclose the circular trajectory of the receiver).
The noise sources are delta-correlated in space and stationary in time, with covariance function $F(t)$.
The wave field is recorded by the moving receiver.
The goal is to understand the relationship between the 
autocorrelation function of the recorded signal and the Green's function between pairs of points along the circular trajectory (see Figure \ref{fig1a}).\\

{\bf The covariance function of the recorded signal.}
The (real-valued) wave field ${u}(t,{\bx})$ emitted by the noise sources satisfies the wave equation 
\begin{equation}
\label{eq:scalar}
 \frac{1}{c^2(\bx)} \frac{\partial^2 {u} }{\partial t^2} -\Delta {u}  =  {s}(t,{\bx})  ,
\end{equation}
where the noise source term is a random process with mean zero and covariance function
$$
\EE \big[ s(t,{\bx})
s(t',{\bx}') \big]  = F(t-t') \delta ({\bx}-{\bx}') \delta_{\partial B} ({\bx}) .
$$
Here $\delta_{\partial B} ({\bx})$ indicates that the covariance is only nonzero on the surface of the ball $B$
and the speed of propagation $c(\bx)$ may be heterogeneous within the ball with center at ${\bf 0}$ and radius $R_1<R_0$ but is homogeneous and equal to $c_0$ outside the ball.
The recorded signal is 
\begin{equation}
\label{def:recordedsignal}
U(t) = u(t ,{\bx}_{\rm r}(t) ) .
\end{equation}
It is recorded over the time interval $[0,2\pi K/v]$, which means that the receiver completes $K$ loops during the recording time window.
We introduce the empirical cross correlation function
\begin{equation}
\label{def:CK}
C_K(\theta,\theta') = \frac{1}{K} \sum_{k=0}^{K-1} U \Big(\frac{\theta+2k\pi}{v} \Big) U \Big(\frac{\theta'+2k\pi}{v} \Big)   ,
\quad \quad \theta, \theta' \in [0,2\pi).
\end{equation}

\begin{proposition}
\label{prop:1}%
When $K \to \infty$, the empirical cross correlation (\ref{def:CK}) converges to the statististical cross correlation
$$
C_K(\theta,\theta')  \stackrel{K \to \infty}{\longrightarrow} C^{(1)}(\theta,\theta')  ,
$$
in probability,
where
\begin{eqnarray}
\nonumber
C^{(1)}(\theta,\theta')&=& 
\EE \Big[ U \Big(\frac{\theta}{v} \Big) U \Big(\frac{\theta'}{v} \Big) \Big]\\
&=&
\frac{1}{2\pi} \int_{-\infty}^\infty  \hat{F}(\omega)  
 \frac{c_0}{\omega} {\rm Im} 
\big\{ \hat{G}(\omega,{\br}_\theta,{\br}_{\theta'}) \big\}
   \exp\Big( -i \frac{\omega}{v}(\theta'-\theta) \Big)   d \omega  ,
   \label{eq:corrstata1}
\end{eqnarray}
${\br}_\theta=(R_0\cos \theta,R_0\sin \theta )$, and $\hat{G}$ is the time-harmonic Green's function solution to 
\begin{equation}
\label{eq:greenhetero}
\Delta_{{\bx}} \hat{G} (\omega,{\bx},{\bx'})+ \frac{\omega^2}{c^2(\bx)} \hat{G}(\omega,{\bx},{\bx'})= - \delta({\bx}-{\bx'})  ,
\end{equation}
with Sommerfeld radiation condition.
\end{proposition}

{\it Proof.}
In the Fourier domain, the (complex-valued) wave field 
$$
\hat{u}(\omega,{\bx})
= \int_{-\infty}^\infty u(t,{\bx}) e^{ i \omega t} dt 
$$
 emitted by the noise sources satisfies the Helmholtz equation 
$$
\Delta  \hat{u} + \frac{\omega^2}{c^2(\bx)} \hat{u}= - \hat{s}(\omega,{\bx})  ,
$$
where the noise source term has the covariance function
$$
\EE \big[ \overline{ \hat{s}(\omega,{\bx})}
\hat{s}(\omega',{\bx}') \big]  = 2 \pi \hat{F}(\omega) \delta(\omega-\omega') 
\delta ({\bx}-{\bx}') \delta_{\partial B} ({\bx}) .
$$
In terms of the Greens' function the wave field is
\begin{equation}
\label{eq:expressuG}
\hat{u}(\omega,{\bx}) = \int_{\RR^2} \hat{G}(\omega,{\bx},{\bx}')
\hat{s}(\omega,{\bx}')
 d{\bx}' .
\end{equation}
The recorded signal (\ref{def:recordedsignal}) is given by 
\begin{eqnarray*}
U(t)
&=& \frac{1}{2\pi} \int_{-\infty}^\infty \hat{u}(\omega,{\bx}_{\rm r}(t)) e^{-i \omega t} d\omega\\
&=& \frac{1}{2\pi} \int_{-\infty}^\infty \int_{\RR^2} \hat{G}(\omega,{\bx}_{\rm r}(t),{\bx}) \hat{s}(\omega,{\bx}) d{\bx} 
e^{-i \omega t} d\omega .
\end{eqnarray*}
We find
$$
\EE \big[ U(t) U(t+\tau) \big]
=
\frac{1}{2\pi} \int_{-\infty}^\infty \hat{F}(\omega) \int_{\partial B} 
 \overline{\hat{G}(\omega,{\bx}_{\rm r}(t),{\bx})} \hat{G}( \omega,{\bx}_{\rm r}(t+\tau),{\bx}) d\sigma({\bx})
e^{-i \omega \tau}   d \omega ,
$$
where $d\sigma({\bx})$ stands for the surface integral.
The covariance function (\ref{def:CK}) can be written in the form
$$
C_K(\theta,\theta') = \frac{1}{K} \sum_{k=0}^{K-1} c_k (\theta,\theta')  ,
$$
where  the random processes $(c_k(\theta,\theta'))_{\theta,\theta'\in [0,2\pi)}$, $k=1,\ldots,K$,
$$
 c_k (\theta,\theta') = U \Big(\frac{\theta+2k\pi}{v} \Big) U \Big(\frac{\theta'+2k\pi}{v} \Big)
$$
are identically distributed and their covariance ${\rm Cov}(  c_k (\theta,\theta') , c_{k'} (\theta,\theta') )$
goes to zero  as $|k-k'| \to \infty$.
As a result 
$$
\EE\big[ \big( C_K(\theta,\theta')  -  \EE \big[  c_1 (\theta,\theta') \big] \big)^2\big] = 
\frac{1}{K^2}\sum_{k,k'=0}^{K-1} {\rm Cov}(  c_k (\theta,\theta') , c_{k'} (\theta,\theta') )  \stackrel{K \to \infty}{\longrightarrow}  0,
$$
and therefore, by Chebyshev's inequality,
$$
C_K(\theta,\theta') \stackrel{K \to \infty}{\longrightarrow}  \EE \big[  c_1 (\theta,\theta') \big] ,
$$
in  probability.
Finally, by Helmholtz-Kirchhoff identity 
(see, for instance \cite[p. 419]{born} or \cite[Theorem 2.33]{ammari})  
we have
\begin{equation}
\label{eq:hk1}
 \int_{\partial B}  \overline{\hat{G}(\omega,{\br}_\theta,{\bx})} \hat{G}( \omega,{\br}_{\theta'},{\bx}) d\sigma({\bx}) = 
\frac{c_0}{\omega} {\rm Im} 
\big\{ \hat{G}(\omega,{\br}_\theta,{\br}_{\theta'}) \big\} ,
\end{equation}
which gives the desired result.
\hfill {\small $\Box$}\\

{\bf Discussion.}
The result presented in Proposition \ref{prop:1} deserves some interpretation. It shows that the autocorrelation function of the recorded signal
is related to the matrix of (the imaginary parts of the) Green's functions between pairs of points along the circular trajectory 
$\big( {\rm Im}  \{ \hat{G}(\omega,{\br}_\theta,{\br}_{\theta'})  \} \big)_{\theta,\theta'\in [0,2\pi)}$.
 However, as shown by (\ref{eq:corrstata1}), only the time component at $(\theta'-\theta)/v$ is accessible.
 To get the full matrix, it is therefore necessary to get the autocorrelation function at different receiver velocities.
 If we assume that we can get the data for all receiver velocities, then  (\ref{eq:corrstata1}) shows that we can get the full matrix.
 If we assume that we can get the data for velocities within the interval $[0,v_{\rm max}]$, then this means 
 that we can get the time components of the Green's function between $\br_\theta$ and $\br_{\theta'}$ 
 within the time interval $[|\theta'-\theta|/v_{\rm max},\infty)$. Since the medium is homogeneous outside the ball $B({\bf 0}, R_1)$, $R_1<R_0$, 
 this means that can can capture the scattered Green's function (i.e. the difference between the full Green's function $\hat{G}$ and the homogeneous Green's function $\hat{G}_0$)
 provided $|\theta'-\theta| \leq 2 (R_1-R_0) v_{\rm max}/c_0$.
 In other words we only have the information related to a diagonal band of the full matrix, whose thickness is limited by the velocity of the receiver.
 In the next section we will address a situation in which the data are collected with a single receiver velocity, but the medium 
 contains only one point-like receiver that can be imaged from the data. 
 
 The result presented in Proposition \ref{prop:1} could be considered as expected. Indeed, 
 the cross correlation function of the signals recorded by two stationary receivers
at  $\br_\theta$ and $\br_{\theta'}$  and emitted by ambient noise sources
  is known to be related to the imaginary part of the Green's function between the two receiver points \cite{garnier16}.
 The standard physical explanation of this result is via an analogy with a time-reversal experiment:
 the cross correlation of the recorded ambient noise signals is the signal recorded by the receiver at $\br_{\theta'}$ 
 during a time-reversal experiment
 in which a short pulse is emitted from $\br_\theta$, recorded by a time-reversal mirror at the surface of the ball $B$,
and remitted, time-reversed,  into the medium.
 However, when the sensors are moving, the analogy with time reversal does not hold anymore as we show in Appendix \ref{app:TR}.
Proposition \ref{prop:1} gives the correct statement when the receiver is moving.\\

{\bf Synthetic experiment.}
It is possible to carry out a simple experiment with one receiver and one source 
to compute synthetically the statistical cross correlation $C^{(1)}$ defined by (\ref{eq:corrstata1}).
Since $F(t)$ is the covariance function of a stationary process, its Fourier transform is nonnegative (by Bochner's theorem).
We define
\begin{equation}
\label{def:synthef}
f(t) = \frac{1}{2 \pi} \int_{-\infty}^\infty  \hat{F}(\omega)^{1/2} e^{-i \omega t} dt.
\end{equation}
The experiment is carried out as follows:\\
1) Record the signal $u(t,{\br}_\theta ; {\bx}_s)$ when the source is at ${\bx}_s \in \partial B$ and
emits the pulse $f(t)$, and  the receiver is stationary at ${\br}_\theta$, $\theta\in [0,2\pi)$.\\
2) Compute the synthetic cross correlation:
$$
C ( \theta,\theta') = \sum_{s=1}^N \int_{-\infty}^\infty u\big(t,{\br}_\theta ; {\bx}_{\rm s}\big) 
u\Big(t+\frac{\theta'-\theta}{v},{\br}_{\theta'} ; {\bx}_{\rm s}\Big) dt ,
$$
when $({\bx}_s)_{s=1}^N$ are the $N$ successive positions of the source that are uniformly
distributed on $\partial B$. Here $v$ is a fixed ``artificial" velocity (in rad/s).

Assuming that the number $N$ is large enough so that we can make the continuum approximation for the sum over $s$,
we can write
$$
C ( \theta,\theta') = \int_{\partial B} \int_{-\infty}^\infty G*f\big(t,{\br}_\theta ; {\bx}\big) 
G*f\Big(t+\frac{\theta' - \theta}{v},{\br}_{\theta'} ; {\bx}\Big) dt d\sigma({\bx}),
$$
up to a multiplicative constant, where $G$ is the time-dependent Green's function and $*$
stands for the convolution product (in $t$).
Therefore we have
$$
C ( \theta,\theta') = \frac{1}{2\pi}
 \int_{-\infty}^\infty |\hat{f}(\omega)|^2 
\frac{c_0}{\omega} {\rm Im} 
\big\{ \hat{G}(\omega,{\br}_\theta,{\br}_{\theta'}) \big\}
 \exp\Big( - i \omega \frac{\theta' -\theta}{v} \Big) d\omega, 
$$
that is to say,
$$
C ( \theta,\theta') = 
C^{(1)} ( \theta,\theta') ,
$$
where $C^{(1)}$ is given by (\ref{eq:corrstata1}),
since $|\hat{f}(\omega)|^2= \hat{F}(\omega)$.

\section{Passive reflector imaging from a receiver moving on a circular trajectory}
\label{sec:ref}%
\begin{figure}
\begin{center}
\begin{tabular}{c}
 \psfig{file=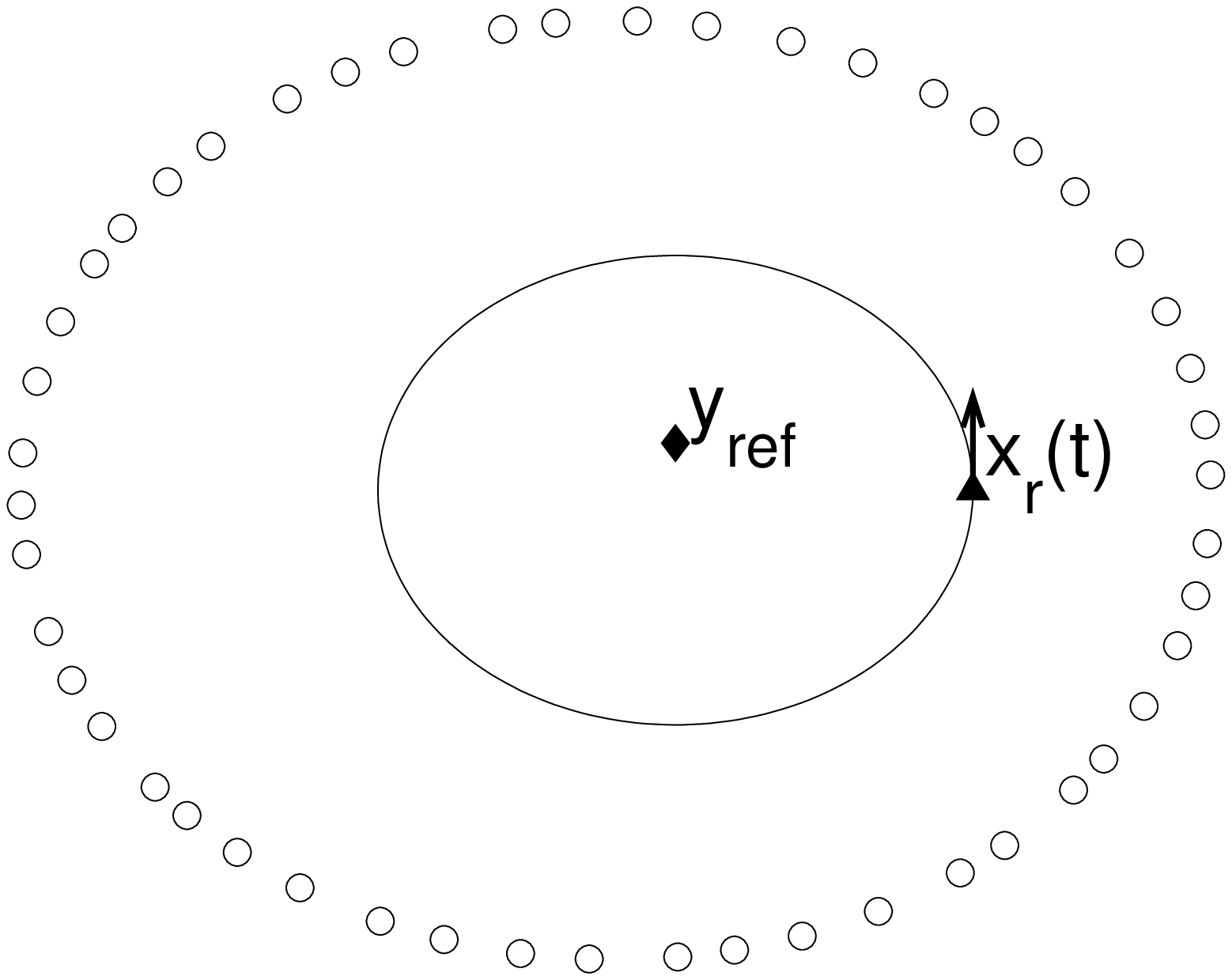,width=6.0cm}
\end{tabular}
\end{center}
\caption{Experimental set-up for passive reflector imaging in Section \ref{sec:ref}. The circles are noise sources (at the surface $\partial B$), 
the triangle is a receiver at ${\bx}_{\rm r}(t)$ on a circular trajectory (with radius $R_0$), and
the diamond is a reflector at ${\by}_{\rm ref}$.
\label{fig1}}
\end{figure}

The set-up is similar to the one addressed in Section~\ref{sec:gre}. The only difference is that 
the complex medium here simply consists of a point-like reflector located at the unknown position ${\by}_{\rm ref}=(x_{\rm ref},y_{\rm ref})$.
In this section the goal is to localize the reflector from the recorded signal  (see Figure \ref{fig1}).

In this section the speed of propagation has the form
$$
 \frac{1}{c^2({\bx})} = \frac{1}{c_0^2} \big( 1 + \nu_{\rm ref} {\bf 1}_{\Omega_{\rm ref}}({\bx} - {\by}_{\rm ref}) \big) .
$$
Here  ${\by}_{\rm ref}$ is the center of the reflector, 
$\Omega_{\rm ref}$ is a small domain that represents the spatial support of the reflector, and $\nu_{\rm ref}$ is the
contrast of the reflector.
In the Born approximation for the reflector the Green's function $\hat{G}$ has the form
\begin{eqnarray}
\label{def:G1a}
\hat{G}(\omega,{\bx},{\bx}') &=&  \hat{G}_0(\omega,{\bx},{\bx}') +
\hat{G}_1(\omega,{\bx},{\bx}') , \\
\hat{G}_1(\omega,{\bx},{\bx}') &=&
\frac{ \omega^2}{c_0^2}\nu_{\rm ref}  \int_{\Omega_{\rm ref} } \hat{G}_0(\omega,{\bx},{\bz} ) \hat{G}_0(\omega,{\bz},{\bx}')d{\bz}  ,
\end{eqnarray}
where the two-dimensional homogeneous Green's function $\hat{G}_0$ is the solution to 
\begin{equation}
\Delta_{{\bx}} \hat{G}_0 (\omega,{\bx},{\bx'})+ \frac{\omega^2}{c^2_0} \hat{G}_0(\omega,{\bx},{\bx'})= - \delta({\bx}-{\bx'})  ,
\end{equation}
with Sommerfeld radiation condition. It is given by
\begin{equation}
\label{def:green0}
\hat{G}_0(\omega,{\bx},{\bx'}) =  \frac{i}{4 } H_0^{(1)} \Big( \frac{\omega}{c_0} |{\bx}-{\bx'}|\Big) ,
\end{equation}
where $H_0^{(1)}$ is the Hankel function of the first kind and of order zero. 
If the reflector can be considered as point-like, then the scattered Green's function $\hat{G}_1$ can be simplified as:
\begin{equation}
\label{def:G1}
\hat{G}_1(\omega,{\bx},{\bx}') =
\frac{\omega^2}{c_0^2} \sigma_{\rm ref} \hat{G}_0(\omega,{\bx},{\by}_{\rm ref}) \hat{G}_0(\omega,{\by}_{\rm ref},{\bx}') ,
\end{equation}
with $\sigma_{\rm ref} = \nu_{\rm ref} |\Omega_{\rm ref}|$, and 
the statistical covariance function $C^{(1)}$ is the sum of two terms:
$$
C^{(1)}(\theta,\theta') = C^{(1)}_0(\theta,\theta') + C^{(1)}_1(\theta,\theta'),
$$
following from the Born approximation of the Green's function.
We study these two contributions in the next two paragraphs.\\

{\bf The direct contribution to the covariance function.}
The direct contribution (i.e. the contribution of the waves that have not been reflected by the reflector)
is 
$$
C^{(1)}_0(\theta,\theta')=\frac{c_0}{2\pi} \int_{-\infty}^\infty \frac{\hat{F}(\omega)}{\omega} {\rm Im} 
\big\{ \hat{G}_0(\omega,{\br}_\theta,{\br}_{\theta'}) \big\}  \exp \Big( -i \frac{\omega}{v}(\theta'-\theta) \Big)  d \omega  ,
$$
with $\hat{G}_0$ the two-dimensional homogeneous Green's function (\ref{def:green0}).
Its imaginary part is 
\begin{equation}
\label{def:green0i}
{\rm Im} \big\{ \hat{G}_0(\omega,{\bx},{\bx'})  \big\} = \frac{1}{4} J_0  \Big( \frac{\omega}{c_0} |{\bx}-{\bx'}|   \Big)
.
\end{equation}
We find
\begin{equation}
C^{(1)}_0 \Big( \theta+\frac{h}{2} ,\theta - \frac{h}{2} \Big) 
= \frac{c_0}{4\pi} \int_{0}^\infty \frac{\hat{F}(\omega)}{\omega}  
 \cos \Big(  \frac{\omega}{v_0} R_0 h \Big) J_0 \Big( 2 \frac{\omega}{c_0} R_0 \big| \sin \big( \frac{h}{2}\big)\big| \Big)
d\omega .
\end{equation}
Using the representation $2\pi J_0(s) = \int_0^{2\pi} e^{i s \sin \phi} d\phi$, we get the two following results
using stationary phase arguments when $\omega_0 R_0/c_0 \gg 1$ (where $\omega_0$ is the central frequency of the sources):\\
1) If $v_0 < c_0$ (i.e. the receiver motion is subsonic), then
there is a unique peak centered at $h=0$, with width $\min \big( c_0/(\omega_0 R_0) , v_0 / ( \omega_{\rm B} R_0)\big)$.
More exactly, under assumption (H1), 
$$
\mbox{(H1)}
 \hspace*{0.35in}
 \begin{array}{l}
 \mbox{the power spectral density is of the form }\\
 \mbox{$\hat{F}(\omega) =\hat{F}_{\rm B}(\omega-\omega_0) + \hat{F}_{\rm B}( \omega+\omega_0)$,}\\
 \mbox{with $\hat{F}_{\rm B}$ even and real and the width $\omega_{\rm B}$ of $\hat{F}_{\rm B}$ is smaller than $\omega_0$,}
 \end{array}
 \hspace*{0.25in}
$$
which also means that $F(t) = e^{- i \omega_0 t }F_{\rm B}(t) +c.c.$ ($c.c.$ stands for complex conjugate), 
and $F_{\rm B}$ is even and real,
we have
$$
C^{(1)}_0 \Big( \theta+\frac{h}{2} ,\theta - \frac{h}{2} \Big) 
= \frac{c_0}{2\omega_0} F_{\rm B} \Big( \frac{R_0}{v_0} h\Big) \cos \Big( \frac{\omega_0}{v_0} R_0 h \Big)
J_0 \Big( \frac{\omega_0}{c_0} R_0 |h| \Big) .
$$
As a function of $h$,
it has the form of a modulated peak centered at $0$, with rapid oscillations at the scale $v_0/(\omega_0 R_0)$, and with radius 
determined by the minimum of the radii of the term in $J_0$ and the term in $F_{\rm B}$.\\
2) If $v_0>c_0$, then there are two other peaks at $h=\pm h_0$, where $h_0\in [0,2\pi]$ is the unique solution to ${\rm sinc}(h_0/2)=c_0/v_0$,
and the widths of these peaks are of the order of the bandwidth  $\omega_{\rm B}$  of the noise sources.
 More exactly,  under assumption (H1), 
for $h$ of the order of $c_0/(\omega_{\rm B} R_0)$,
we have
\begin{eqnarray*}
C^{(1)}_0 \Big( \theta+\frac{h_0+h}{2} ,\theta - \frac{h_0+h}{2} \Big) 
= \frac{\sqrt{v_0}}{\sqrt{ 2\pi R_0 h_0 \omega_0} }\frac{c_0}{2\omega_0} 
\cos \Big( \omega_0 \frac{R_0}{c_0} \big( \frac{c_0}{v_0} -\cos \frac{h_0}{2}   \big) h+\frac{\pi}{4} \Big) \\
\times
F_{\rm B} \Big(\frac{R_0}{c_0} \big( \frac{c_0}{v_0} - \cos \frac{h_0}{2}\big) h \Big) .
\end{eqnarray*}
The amplitudes of these secondary peaks are smaller than the main peak centered at $0$
(with a ratio in the amplitudes of the order of $\sqrt{v_0/(\omega_0 R_0)}$), 
and their widths are larger (with a ratio in the widths of the order of $\omega_0 /\omega_{\rm B}$).\\

{\bf The scattered contribution to the covariance function.}
The scattered contribution (i.e. the contribution of the waves that have been reflected by the reflector)
is 
$$
C^{(1)}_1(\theta,\theta')=\frac{c_0}{2\pi} \int_{-\infty}^\infty \frac{\hat{F}(\omega)}{\omega} {\rm Im} 
\big\{ \hat{G}_1(\omega,{\br}_\theta,{\br}_{\theta'}) \big\}  \exp \Big( -i \frac{\omega}{v}(\theta'-\theta) \Big) d \omega  ,
$$
with $\hat{G}_1$  given by (\ref{def:G1}).
If the distance from the reflector to the sphere with radius $R_0$ is larger than the typical wavelength $\lambda_0 = 2\pi c_0/\omega_0$,
we can use the asymptotic form of the two-dimensional homogeneous Green's function 
based on the expansion (\ref{eq:asymptHankel}) of the Hankel function and we get
\begin{eqnarray}
\nonumber
&& C^{(1)}_1\Big( \theta+\frac{h}{2} ,\theta - \frac{h}{2} \Big) = \frac{\sigma_{\rm ref}}{32\pi^2} 
\int_{-\infty}^\infty \frac{ \hat{F}(\omega) }{|{\br}_{\theta+\frac{h}{2} }-{\by}_{\rm ref}|^{1/2} \, |{\by}_{\rm ref}-{\br}_{\theta-\frac{h}{2} }|^{1/2}}
\\
\nonumber
&& \hspace*{0.2in} \times
\Big[ 
\exp\Big(  i \frac{\omega}{c_0} ( |{\br}_{\theta+\frac{h}{2} }-{\by}_{\rm ref}| + |{\by}_{\rm ref}-{\br}_{\theta-\frac{h}{2} }| )\Big) \\
&& \hspace*{0.3in} 
+
\exp\Big( -i \frac{\omega}{c_0} ( |{\br}_{ \theta+\frac{h}{2} }-{\by}_{\rm ref}| + |{\by}_{\rm ref}-{\br}_{\theta- \frac{h}{2} }| )\Big)
\Big] 
\exp \Big( i \frac{\omega}{v} h\Big)   d \omega  .
\label{eq:expC11v1b}
\end{eqnarray}

We can simplify this expression under different conditions, as shown in the next lemma.

\begin{lemma}
\label{lem:imag1}%
\begin{enumerate}
\item
When $v_0  \ll c_0$, we have for any $r$ of the order of $R_0$:
 \begin{eqnarray}
\nonumber
C^{(1)}_1 \Big( \theta+\frac{v r}{2c_0} ,\theta - \frac{v r}{2c_0} \Big) &=&
\frac{\sigma_{\rm ref}}{16\pi R_0} F \Big( \frac{r }{c_0} -2 \frac{|{\br}_\theta -{\by}_{\rm ref}|}{c_0} 
 \Big) 
\\
&&
 + \frac{\sigma_{\rm ref}}{16\pi R_0} F \Big( \frac{r}{c_0} +2 \frac{|{\br}_\theta -{\by}_{\rm ref}|}{c_0} 
\Big) .
\label{eq:expC11v2}
\end{eqnarray}
\item
When $v_0 \lesssim  c_0$ and $R_0 \gg |{\by}_{\rm ref}|$, we have: 
\begin{eqnarray}
\nonumber
&&
C^{(1)}_1 \Big( \theta+\frac{h}{2} ,\theta - \frac{h}{2} \Big) \\
\nonumber
&&=
\frac{\sigma_{\rm ref}}{16\pi R_0} F \Big( \frac{h}{v} -2 \frac{|{\br}_\theta -{\by}_{\rm ref}|}{c_0} - 4 \sin^2 \big( \frac{h}{4}\big)
\frac{ x_{\rm ref}\cos \theta  + y_{\rm ref} \sin \theta}{c_0} 
 \Big) \\
&&\quad   + \frac{\sigma_{\rm ref}}{16\pi R_0} F \Big( \frac{h}{v} +2 \frac{|{\br}_\theta -{\by}_{\rm ref}|}{c_0} +4 \sin^2 \big( \frac{h}{4}\big)
\frac{x_{\rm ref}\cos \theta  + y_{\rm ref} \sin \theta}{c_0} 
\Big) .
\label{eq:expC11v1}
\end{eqnarray}
\end{enumerate}
\end{lemma}
Obviously the covariance function contains information about the reflector position that can be extracted by migration,
as shown in the next paragraph.

{\it Proof.}
When $v_0  \ll c_0$, we have for any $r$ of the order of $R_0$:
$$
|{\br}_{\theta+\frac{v r}{2c_0} }-{\by}_{\rm ref}|  = 
 |{\br}_\theta-{\by}_{\rm ref}|  + O \Big( \frac{v_0}{c_0} \Big).
$$
By substitution into (\ref{eq:expC11v1b}) we find (\ref{eq:expC11v2}).\\
- When $v_0 \lesssim  c_0$ and $R_0 \gg |{\by}_{\rm ref}|$, we have for $k \geq 1$: 
$$
\frac{\partial^{2k}}{\partial \theta^{2k}}  |{\br}_\theta-{\by}_{\rm ref}| = -  (-1)^{k} \big(x_{\rm ref} \cos \theta
+ y_{\rm ref} \sin \theta \big)  
 + O \Big( \frac{|{\by}_{\rm ref}|^2}{R_0} \Big)  .
$$
From the expansion valid for any $h$
$$
|{\br}_{\theta+h/2}-{\by}_{\rm ref}| + |{\by}_{\rm ref}-{\br}_{\theta-h/2}| = 
2 |{\br}_\theta-{\by}_{\rm ref}|  +
2 \sum_{k=1}^\infty \frac{1}{(2k) !} \Big( \frac{h}{2} \Big)^{2k} \frac{\partial^{2k}}{\partial \theta^{2k}}  |{\br}_\theta-{\by}_{\rm ref}| ,
$$
we get
\begin{eqnarray*}
&& |{\br}_{\theta+h/2}-{\by}_{\rm ref}| + |{\by}_{\rm ref}-{\br}_{\theta-h/2}| \\
&&= 
2 |{\br}_\theta-{\by}_{\rm ref}|  -
2 \sum_{k=1}^\infty \frac{(-1)^k}{(2k) !} \Big( \frac{h}{2} \Big)^{2k}  \big(x_{\rm ref} \cos \theta
+ y_{\rm ref} \sin \theta \big)  
 + O \Big( \frac{|{\by}_{\rm ref}|^2}{R_0} \Big) 
\\
&&=2 |{\br}_\theta-{\by}_{\rm ref}|  +
4  \sin^2 \Big( \frac{h}{4} \Big)   \big(x_{\rm ref} \cos \theta
+ y_{\rm ref} \sin \theta \big)  
 + O \Big( \frac{|{\by}_{\rm ref}|^2}{R_0} \Big) 
.
\end{eqnarray*}
Here we have used the fact that $\cos (s) = \sum_{k \geq 0} (-1)^k  s^{2k} / [(2k)!]$ and $1-\cos(s)=2\sin^2(s/2)$.
Therefore we find (\ref{eq:expC11v1}).
\hfill {\small $\Box$}\\

{\bf The imaging function.}
Motivated by Lemma \ref{lem:imag1} that exhibits the presence of a peak in the 
cross correlation $C_K(\theta+h,\theta-h)$ at $h= ({v}/{c_0}) \big|{\br}_\theta-{\by}_{\rm ref}\big|$,
we first propose to image the reflector with the imaging function defined by
\begin{equation}
\label{def:imagfunction1}
{\mathcal I}( {\by}^S) = \int_0^{2\pi}  C_K\Big(\theta + \frac{v}{c_0} \big|{\br}_\theta-{\by}^S\big| ,
\theta-\frac{v}{c_0} \big|{\br}_\theta-{\by}^S \big| \Big)  d\theta   .
\end{equation}
The covariance function $C_K$ contains the direct and scattered contributions  analyzed here above. The direct contribution 
does not give any peak in the imaging function (\ref{def:imagfunction1}) while the scattered contribution gives a peak.

When $v_0  \ll c_0$, we find using (\ref{eq:expC11v2})  that
\begin{equation}
{\mathcal I}( {\by}^S) = 
\frac{\sigma_{\rm ref}}{8 \pi R_0} \int_0^\infty \hat{F}(\omega) J_0 \Big( 2 \frac{\omega}{c_0} \big|{\by}^S -  {\by}_{\rm ref} \big|\Big) d\omega  .
\end{equation}

When $v_0\lesssim c_0$ and $R_0 \gg  |{\by}_{\rm ref}|,  |{\by}^S|$, we find using (\ref{eq:expC11v1}) that
\begin{equation}
\label{def:imagfunction1:3}
{\mathcal I}( {\by}^S) = 
\frac{\sigma_{\rm ref}}{8 \pi R_0} \int_0^\infty \hat{F}(\omega) J_0 \Big( 2 \frac{\omega}{c_0} \big|{\by}^S - \cos (\frac{v_0}{c_0}) {\by}_{\rm ref} \big|\Big) d\omega  .
\end{equation}
This expression is correct provided $({v_0^2}/{c_0^2}) [ {|{\by}_{\rm ref}|^2}/({R_0 \lambda}) ] \ll 1$
where  $\lambda$ is the typical wavelength. We study the corrective term when this condition
is not fulfilled in Appendix \ref{app:B}.\\
The expression (\ref{def:imagfunction1:3}) shows that
 imaging function (\ref{def:imagfunction1}) has a peak with width given by $c_0/(2\omega_0)$ 
(where $\omega_0$ is the central frequency of the noise sources)
and centered not on the exact location of the reflector,
but on the position $\cos ({v_0}/{c_0}) {\by}_{\rm ref}$. In other words, the imaging function 
(\ref{def:imagfunction1}) plots an image of the medium rescaled by the factor $\cos ({v_0}/{c_0})$.
This rescaling is due to the Doppler effect.
Therefore we can propose a rescaled version of the imaging function:
\begin{equation}
\widetilde{\mathcal I}( {\by}^S) = \int_0^{2\pi}  C_K
\Big(
\theta +  \frac{v}{c_0} \big|{\br}_\theta-  \cos(\frac{v_0}{c_0}) {\by}^S\big| 
,
\theta - \frac{v}{c_0} \big| {\br}_\theta- \cos(\frac{v_0}{c_0})  {\by}^S\big| 
\Big)  
d\theta  .
\end{equation}
We find, when $v_0\lesssim c_0$ and $R_0 \gg  |{\by}_{\rm ref}|,  |{\by}^S|$, that
\begin{equation}
\label{eq:modimag1}
\widetilde{\mathcal I}( {\by}^S) = 
\frac{\sigma_{\rm ref}}{8 \pi R_0} \int_0^\infty \hat{F}(\omega) J_0 \Big( 2 \frac{\omega}{c_0} \cos(\frac{v_0}{c_0}) 
\big|{\by}^S - {\by}_{\rm ref} \big|\Big) d\omega  .
\end{equation}
The rescaled imaging function has a peak centered at the location of the reflector
with width given by $2.4 \lambda_0/ [ 2\pi \cos({v_0}/{c_0}) ]$,
where $\lambda_0 = 2\pi c_0/\omega_0$ is the central wavelength (and $2.4$ is approximately the first zero of 
the Bessel function $J_0$).
Note that  resolution is reduced when the velocity  of the receiver increases. 
This can be interpreted as a consequence of Doppler effect.

Note that, in order to compute the imaging function,
it is not required to evaluate and store $C_K(\theta,\theta')$ for all $\theta,\theta'\in [0,2\pi)$.
It is sufficient to compute it for a narrow band along the diagonal $\theta'=\theta$, the width of the diagonal band being 
$2 v_0/c_0$.
Note also that the direct contribution of the covariance function does not give any peak in the imaging function
(\ref{def:imagfunction1}) but it may give an incoherent background in the image. Therefore the peak due to the scattered contribution
can be visible provided the scattering coefficient $\sigma_{\rm ref}$ of the reflector is not too small.

\section{Passive reflector imaging from a receiver moving on a linear trajectory}
\label{sec:ref2}%

The goal of this section is to show that the results obtained in Section \ref{sec:ref}
are not specific to the case where the receiver moves along a circular trajectory.
Here we extend the result to the case of a linear trajectory.
This does not change qualitatively the picture but this 
affects quantitatively the resolution properties of the corresponding imaging function.
We can anticipate that such results could be obtained for other configurations.\\

\begin{figure}
\begin{center}
\begin{tabular}{c}
 \psfig{file=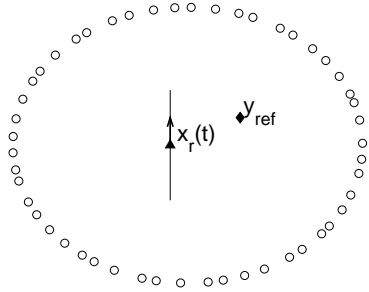,width=6.0cm}
\end{tabular}
\end{center}
\caption{Experimental set-up for passive reflector imaging in Section \ref{sec:ref2}. 
The circles are noise sources (at the surface $\partial B$), 
the triangle is a receiver on a linear trajectory (with length $a$), and
the diamond is a reflector.
\label{fig3}}
\end{figure}

{\bf Experimental set-up.}
We consider a moving receiver.
Its position is ${\bx}_{\rm r}(t) = (v_0 t, 0)$, for $t \in [-a/(2v_0), a/(2v_0)]$, where $v_0$ is its velocity
and $a$ is the length of its linear trajectory.
Ambient noise sources located at the surface of a large ball $B$ emit stationary random signals
(the ball $B$ does not need to be centered at ${\bf 0}$, but it needs to enclose the trajectory of the receiver).
The noise sources are delta-correlated in space and stationary in time, with covariance function $F(t)$.
The wave field is recorded by the moving receiver.
The goal is to image from the recorded signal a point-like reflector located at ${\by}_{\rm ref}=(x_{\rm ref},y_{\rm ref})$
(see Figure \ref{fig3}).\\ 
We repeat $K$ times the experiment, that is to say we record 
$K$ times the signal received by the sensor ${\bx}_{\rm r}(t)$, $k=1,\ldots,K$,
with $K$ independent realizations of the signals $n^{(k)}(t,{\bx})$ emitted by the noise sources. This is necessary to achieve statistical stability
(i.e. the empirical cross correlation is approximately equal to the statistical cross correlation).\\

{\bf The covariance function of the recorded signal.}
The recorded signal during the $k$-th experiment  is 
\begin{equation}
U^{(k)}(t) = u^{(k)} (t ,{\bx}_{\rm r}(t) ) , 
\end{equation}
with
\begin{equation}
u^{(k)}(t,{\bx}) = \int_{-\infty}^\infty \int_{\RR^2} G(t-s,{\bx},{\bx}') n^{(k)}(s,{\bx}') d{\bx}' ds .
\end{equation}
We introduce the empirical covariance function 
\begin{equation}
\label{def:CKbis}
C_K(x,x') = \frac{1}{K} \sum_{k=1}^{K} U^{(k)} \Big(\frac{x}{v_0} \Big) U^{(k)} \Big(\frac{x'}{v_0} \Big)   ,
\quad \quad x, x'\in (-a/2,a/2).
\end{equation}
We can proceed as in the previous section to get the following result.

\begin{proposition}
When $K \to \infty$ the empirical  covariance function converges to the statistical cross correlation
$$
C_K(x,x')  \stackrel{K \to \infty}{\longrightarrow} C^{(1)}(x,x') ,
$$
in probability,
where
\begin{eqnarray}
\nonumber
C^{(1)}(x,x')&=& 
\EE \Big[ U^{(1)} \Big(\frac{x}{v_0} \Big)  U^{(1)} \Big(\frac{x'}{v_0} \Big)  \Big]\\
&=&
\frac{1}{2\pi} \int_{-\infty}^\infty  \hat{F}(\omega)  
 \frac{c_0}{\omega} {\rm Im} 
\big\{ \hat{G}(\omega,{\br}_x,{\br}_{x'}) \big\}
  \exp \Big( -i \frac{\omega}{v_0}(x'-x) \Big)  d \omega  ,
  \label{eq:corrstata1bis}
\end{eqnarray}
where ${\br}_x=(x,0)$  and $\hat{G}$ is the Green's function in the presence of the reflector at ${\by}_{\rm ref}$.
\end{proposition}

The statistical covariance function $C^{(1)}$ can then be decomposed into the sum of two terms
following from the Born approximation of the Green's function.
We study these two contributions in the next two paragraphs.\\

{\bf The direct contribution to the covariance function.}
The direct contribution (i.e. the contribution of the waves that have not been reflected by the reflector)
is 
$$
C^{(1)}_0(x,x')=\frac{c_0}{2\pi} \int_{-\infty}^\infty \frac{\hat{F}(\omega)}{\omega} {\rm Im} 
\big\{ \hat{G}_0(\omega,{\br}_x,{\br}_{x'}) \big\} \exp\Big( -i \frac{\omega}{v_0}(x'-x) \Big)   d \omega  ,
$$
with $\hat{G}_0$ the two-dimensional homogeneous Green's function (\ref{def:green0}).
We find
\begin{equation}
C^{(1)}_0 \Big( X+\frac{\rho}{2} ,X - \frac{\rho}{2} \Big) 
= \frac{c_0}{4\pi} \int_{0}^\infty \frac{\hat{F}(\omega)}{\omega}  
\cos \Big( \frac{\omega}{v_0} \rho \Big) J_0 \Big( \frac{\omega}{c_0}  |\rho| \Big)
d\omega .
\end{equation}
Using the representation $2\pi J_0(s) = \int_0^{2\pi} e^{i s \sin \phi} d\phi$
and stationary phase arguments when
 $v_0/c_0 \neq 1$, we find that
there is a unique peak centered at $\rho=0$, with width $\min( c_0/\omega_0 , v_0/\omega_{\rm B}) $
(where $\omega_0$ and $\omega_{\rm B}$ are the central frequency and bandwidth of the sources).
More exactly, under assumption (H1), we have
$$
C^{(1)}_0  \Big( X+\frac{\rho}{2} ,X - \frac{\rho}{2} \Big) 
= \frac{c_0}{2 \omega_0} F_{\rm B}\Big( \frac{\rho}{v_0}\Big) \cos \Big( \frac{\omega_0}{v_0} \rho \Big)
J_0 \Big( \frac{\omega_0}{c_0} |\rho| \Big) .
$$

{\bf The scattered contribution to the covariance function.}
The scattered contribution (i.e. the contribution of the waves that have been reflected by the reflector)
is 
$$
C^{(1)}_1(x,x')=\frac{c_0}{2\pi} \int_{-\infty}^\infty \frac{\hat{F}(\omega)}{\omega} {\rm Im} 
\big\{ \hat{G}_1(\omega,{\br}_x,{\br}_{x'}) \big\}  \exp\Big( -i \frac{\omega}{v_0}(x'-x) \Big)  d \omega  ,
$$
with $\hat{G}_1$  given by (\ref{def:G1}).
If the distance from the reflector to the linear trajectory is larger than the typical wavelength,
we can use the asymptotic form of the two-dimensional homogeneous Green's function 
based on the expansion (\ref{eq:asymptHankel}) of the Hankel function and we get
\begin{eqnarray*}
&&
C^{(1)}_1\Big( X+\frac{\rho}{2} ,X - \frac{\rho}{2} \Big) =\frac{\sigma_{\rm ref}}{32\pi^2} 
\int_{-\infty}^\infty \frac{ \hat{F}(\omega) }{|{\br}_{X+\frac{\rho}{2} }-{\by}_{\rm ref}|^{1/2} \, |{\by}_{\rm ref}-{\br}_{X-\frac{\rho}{2} }|^{1/2}}
\\
&&\hspace*{0.2in}
\times
\Big[ 
\exp \Big( i \frac{\omega}{c_0} ( |{\br}_{X+\frac{\rho}{2} }-{\by}_{\rm ref}| + |{\by}_{\rm ref}-{\br}_{X-\frac{\rho}{2} }| ) \Big) \\
&&\hspace*{0.3in}
+
\exp \Big( -i \frac{\omega}{c_0} ( |{\br}_{X+\frac{\rho}{2} }-{\by}_{\rm ref}| + |{\by}_{\rm ref}-{\br}_{X- \frac{\rho}{2} }| ) \Big)
\Big] 
\exp\Big( i \frac{\omega}{v_0} \rho\Big)  d \omega  .
\end{eqnarray*}
If $\rho$ is small, then we can expand
\begin{eqnarray*}
&& C^{(1)}_1\Big( X+\frac{\rho}{2} ,X - \frac{\rho}{2} \Big) = \frac{\sigma_{\rm ref}}{32\pi^2} 
\int_{-\infty}^\infty \frac{ \hat{F}(\omega) }{|{\br}_{X }-{\by}_{\rm ref}| }
\\
&&\hspace*{0.2in}\times
\Big[ 
\exp\Big(  i \frac{\omega}{c_0} (2 |{\br}_{X }-{\by}_{\rm ref}| + \frac{y_{\rm ref}^2}{|{\by}_{\rm ref}-{\br}_{X}|^3} \frac{\rho^2}{4} )\Big) \\
&&\hspace*{0.3in}
+
\exp\Big(  -i \frac{\omega}{c_0} (2 |{\br}_{X }-{\by}_{\rm ref}| + \frac{y_{\rm ref}^2}{|{\by}_{\rm ref}-{\br}_{X}|^3} \frac{\rho^2}{4} )\Big)
\Big] 
\exp\Big( i \frac{\omega}{v_0} \rho\Big)   d \omega  ,
\end{eqnarray*}
with the notation  ${\by}_{\rm ref}=(x_{\rm ref},y_{\rm ref})$.\\

{\bf The imaging function.}
We propose to image the reflector in the subsonic regime $v_0<c_0$ with the imaging function defined by
\begin{equation}
\label{def:imagfunction1bis}
{\mathcal I}( {\by}^S) = \frac{1}{a} \int_{-a/2}^{a/2}   \big|{\br}_X-{\by}^S \big| C_K\Big(X + \frac{v_0}{c_0} \big|{\br}_X-{\by}^S\big| ,
X-\frac{v_0}{c_0} \big|{\br}_X-{\by}^S \big| \Big)  dX   .
\end{equation}
This imaging function is a weighted migration function, and the choice of the weight $\big|{\br}_X-{\by}^S \big|$
is justified by the forthcoming analysis that shows that this weight compensates for the geometric decay  
of the product of the two Green's function contained in the cross correlation.
We analyze this imaging function when the reflector is located far from the linear trajectory 
in the sense that ${\by}_{\rm ref}=(x_{\rm ref},y_{\rm ref})$ with $x_{\rm ref} \sim a \ll y_{\rm ref}$.
Then, parameterizing 
$$
{\by}^S={\by}_{\rm ref}+(\xi,\eta), 
$$
we find
\begin{eqnarray*}
{\mathcal I}( {\by}^S) = \frac{\sigma_{\rm ref}}{32 \pi^2 a}
 \int_{-\infty}^\infty \hat{F}(\omega)  \int_{-a/2}^{a/2}
\exp\Big(  -i \frac{\omega}{c_0} \big[ 2 \big(1-\frac{v_0^2}{c_0^2}\big)\big( \frac{(X-x_{\rm ref})\xi}{y_{\rm ref}} - \eta \big) \\
+\frac{v_0^2 y_{\rm ref}}{c_0^2} - \frac{v_0^2 (X-x_{\rm ref})^2}{2c_0^2 y_{\rm ref}} \big] \Big)
dX d\omega  .
\end{eqnarray*}
If, additionally, assumption (H1) holds, then 
\begin{eqnarray*}
&& \hspace*{-0.25in} {\mathcal I}( {\by}^S) \\
&& \hspace*{-0.25in} =  \frac{\sigma_{\rm ref}}{32 \pi^2} 
\Big[\frac{1}{a} 
\int_{-a/2}^{a/2} 
\exp \Big(- 2 i \frac{\omega_0}{c_0} \big(1-\frac{v_0^2}{c_0^2}\big) \frac{X}{y_{\rm ref}} \big( \xi+ \frac{\frac{v_0^2}{2c_0^2}}{1-\frac{v_0^2}{c_0^2}} x_{\rm ref} \big)  + i \frac{\omega_0}{c_0} \frac{v_0^2}{c_0^2} \frac{X^2}{2 y_{\rm ref}}\Big) dX \Big] \\
&&  \hspace*{-0.15in} \times 
\exp \Big(
2 i \frac{\omega_0}{c_0} \big(1-\frac{v_0^2}{c_0^2}\big)\big( \eta  -
 \frac{\frac{v_0^2}{2c_0^2}}{1-\frac{v_0^2}{c_0^2}} y_{\rm ref} \big)
+
2 i \frac{\omega_0}{c_0} \big(1-\frac{v_0^2}{c_0^2}\big)
\big( \xi+ \frac{\frac{v_0^2}{4c_0^2}}{1-\frac{v_0^2}{c_0^2}} x_{\rm ref} \big) \frac{x_{\rm ref}}{y_{\rm ref}}
  \Big)
 \\
&& \hspace*{-0.15in} \times
\Big[
 \int \hat{F}_{\rm B}(\omega) 
\exp\Big( 2i \frac{\omega}{c_0}  \big(1-\frac{v_0^2}{c_0^2}\big) \big[ \big( \eta  -
 \frac{\frac{v_0^2}{2c_0^2}}{1-\frac{v_0^2}{c_0^2}} y_{\rm ref} \big)
 +
 \big( \xi+ \frac{\frac{v_0^2}{4c_0^2}}{1-\frac{v_0^2}{c_0^2}} x_{\rm ref} \big) \frac{x_{\rm ref}}{y_{\rm ref}}
 \big]
  \Big) d\omega \Big] \\
&& \hspace*{-0.15in} + c.c.
\end{eqnarray*}
When $v_0 \ll c_0$, we have 
$$
{\mathcal I}( {\by}^S) = \frac{\sigma_{\rm ref}}{8 \pi} {\rm sinc} \Big( \frac{\omega_0 a \xi}{y_{\rm ref}} \Big)
\cos \Big(2 \frac{\omega_0}{c_0}  
(\eta +\frac{x_{\rm ref}}{y_{\rm ref}} \xi) 
\Big) F_{\rm B} \Big( \frac{2}{c_0}(\eta +\frac{x_{\rm ref}}{y_{\rm ref}} \xi)  \Big)  .
$$
This shows that the cross range resolution is $\lambda_0 y_{\rm ref} /a$ and the range resolution is $c_0/\omega_{\rm B}$,
where $\lambda_0=2\pi c_0/\omega_0$ is the central wavelength.
These resolution formulas are similar to the case of a passive sensor array extending along the line $[-a,a]\times\{0\}$ \cite{garnier10}.
If the velocity $v_0$ is not negligible compared to $c_0$, then
the image is shifted and slightly blurred (blurring happens when $ ({\omega_0 a^2})/({c_0 y_{\rm ref}}) > {c_0^2}/{v_0^2}$).
It is possible to mitigate -at least partly- these effects by using the modified imaging function:
\begin{eqnarray}
\label{def:imagfunction1tildebis}
&&\widetilde{\mathcal I}( {\by}^S) = \frac{1}{a} \int_{-a/2}^{a/2}   \big|{\br}_{X} - {\by}^S \big| 
C_K\big(X + \Delta(X)  , 
X-\Delta(X)  \big)  dX   ,\\
\nonumber
&& \mbox{where } \Delta(X) =\frac{v_0}{c_0}\big|{\br}_{X}-{\by}^S\big| - \frac{v_0^3}{2c_0^3}  \frac{{y^S}^2}{\big|{\br}_{X}-{\by}^S\big|} ,
\end{eqnarray}
with the notation ${\by}^S=(x^S,y^S)$.
We then find (keeping terms of order $1$, $v_0/c_0$, and $v_0^2/c_0^2$):
\begin{eqnarray*}
\widetilde{\mathcal I}( {\by}^S) &=& \frac{\sigma_{\rm ref}}{32 \pi^2 a}
 \int_{-\infty}^\infty \hat{F}(\omega) \\
 &&\times \int_{-a/2}^{a/2}
\exp \Big( 2 i \frac{\omega}{c_0}   \big(1-\frac{3v_0^2}{2c_0^2}\big)  \frac{(X-x_{\rm ref})\xi}{y_{\rm ref}} 
-2 i \frac{\omega}{c_0}   \big(1-\frac{v_0^2}{2c_0^2}\big)  \eta   \Big)
dX d\omega  .
\end{eqnarray*}
If, additionally,
assumption (H1) holds,  then
\begin{eqnarray*}
\widetilde{\mathcal I}( {\by}^S) = \frac{\sigma_{\rm ref}}{8 \pi} {\rm sinc} \Big(  \big(1-\frac{3v_0^2}{2c_0^2}\big)\frac{\omega_0 a \xi}{y_{\rm ref}} \Big)
\cos \Big(2  \big(1-\frac{v_0^2}{2c_0^2}\big)\frac{\omega_0}{c_0}\big( \eta +
\frac{1-\frac{3v_0^2}{2c_0^2}}{1-\frac{v_0^2}{2c_0^2}}\frac{x_{\rm ref}}{y_{\rm ref}} \xi \big)  
\Big) \\
\times
 F_{\rm B} \Big( \frac{2}{c_0} \big(1-\frac{v_0^2}{2c_0^2}\big) \big( \eta +
\frac{1-\frac{3v_0^2}{2c_0^2}}{1-\frac{v_0^2}{2c_0^2}}\frac{x_{\rm ref}}{y_{\rm ref}} \xi \big) \Big)  .
\end{eqnarray*}
This shows that the modified imaging function gives the right position of the reflector,
at least up to terms of order two in $v_0/c_0$.
Note, however, that both cross range and range resolution are reduced when $v_0$ increases:
the cross range resolution is $\lambda_0 y_{\rm ref} /\{a[1-3v_0^2/(2c_0^2)]\}$ and the range resolution is 
$c_0/\{\omega_{\rm B} [1-v_0^2/(2c_0^2)]\}$. 
By comparing with (\ref{eq:modimag1}) we can see that the relative reductions in resolution 
are of the same order in the case of a circular trajectory and in the case of a linear trajectory.
The cross-range resolution turns out to be relatively more affected, and this can 
be explained by the fact that this is the direction of the motion.
\\

{\bf Synthetic experiment.}
It is possible to carry out a simple experiment with one receiver and one source 
to compute synthetically the statistical cross correlation $C^{(1)}$ defined by (\ref{eq:corrstata1bis}).
We define the pulse profile $f(t)$ as in (\ref{def:synthef}).
The experiment is carried out as follows:\\
1) Record the signal $u(t,{\br}_x ; {\bx}_s)$ when the source is at ${\bx}_s \in \partial B$ and
emits $f(t)$, while the receiver is at ${\br}_x=(x,0)$, $x\in (-a/2,a/2)$.\\
2) Compute the synthetic cross correlation:
$$
C (x,x') = \sum_{s=1}^N \int_{-\infty}^\infty u\big(t,{\br}_x ; {\bx}_s\big) 
u\Big(t+\frac{x'-x}{v_0},{\br}_{x'} ; {\bx}_s\Big) dt ,
$$
when $({\bx}_s)_{s=1}^N$ are the $N$ successive positions of the source that are uniformly
distributed on $\partial B$. Here $v_0$ is a fixed ``artificial" velocity.
Assuming that the number $N$ is large enough so that we can make the continuum approximation for the sum over $s$,
we can show as in Section \ref{sec:gre} that (up to a multiplicative constant)
$$
C ( x,x') = 
C^{(1)} (x,x') ,
$$
where $C^{(1)}$ is given by (\ref{eq:corrstata1bis}).

\section{Green's function estimation with a noise source moving on a circular trajectory}
\label{sec:1}%
In this section we consider another situation related to passive Green's function estimation, 
when the sources themselves are moving. The  analysis of the previous sections can be extended to this 
situation and we show the surprising result that a unique point-like source emitting a stationary random signal
and moving along a periodic trajectory provides an illumination that is appropriate for passive
Green's function estimation from the signals recorded by two receivers.
As an application we will show that travel time estimation between two receivers can be carried out by using the signals 
emitted by the moving point source and recorded by the receivers.\\

\begin{figure}
\begin{center}
\begin{tabular}{c}
 \psfig{file=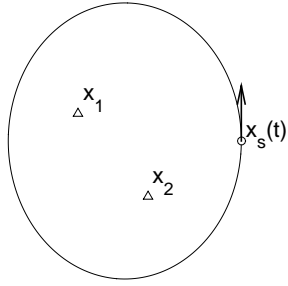,width=6.0cm}
\end{tabular}
\end{center}
\caption{Experimental set-up for passive Green's function estimation in Section \ref{sec:1}. The circle is the trajectory of the moving source ${\bx}_{\rm s}(t)$
and the two triangles are two observation points at $\bx_1$ and $\bx_2$.
\label{fig0}}
\end{figure}

{\bf Experimental set-up.}
We consider a moving point-like source emitting a stationary random signal $f(t)$ with mean zero and
covariance function $\EE[f(t) f(t')]=F(t-t')$.
Its position is ${\bx}_{\rm s}(t) = (R_0 \cos(vt), R_0 \sin(vt))$, where $R_0>0$ is the radius
of the circular trajectory of the source and $v$ is its angular velocity (its linear  velocity is $v_0=vR_0$).
The signals are recorded at two points ${\bx_1}$ and ${\bx_2}$ within the ball with radius $R_0$ (see Figure \ref{fig0}).
Our goal is to express the cross correlation of the signals recorded by the two receivers
in terms of the Green's function between them 
and to clarify the effect of the velocity of the source.\\

{\bf Covariance function of the recorded signals.}
The wave field ${u}(t,{\bx})$ emitted by the moving source satisfies the scalar wave equation (\ref{eq:scalar}) 
where the source term is  
\begin{equation}
s(t,{\bx}) = \delta \big( {\bx}-{\bx}_{\rm s}(t)\big) f(t)  ,
\end{equation}
and $c(\bx)$ is the speed of propagation of the medium, that may be heterogeneous, but that is assumed to be homogeneous with velocity $c_0$
outside the ball with center at ${\bf 0}$ and radius $R_1<R_0$.
The empirical covariance function of the signals recorded at ${\bx_1}$ and ${\bx_2}$ is defined by
\begin{equation}
\label{def:CT0}
C_T(\tau) = \frac{1}{T} \int_0^T u(t,{\bx_1}) u(t+\tau,{\bx_2}) dt .
\end{equation}
The following proposition describes the convergence of the empirical covariance function (\ref{def:CT0}) towards the statistical cross correlation (\ref{eq:C1aprop1})
or (\ref{eq:C1bprop1}) as the recording time increases. 
The relation between the statistical cross correlation and the (imaginary part of the) Green's function will be clarified in Proposition \ref{prop:0}.

\begin{proposition}
When $vT$ goes to infinity, the empirical cross correlation converges to the statististical cross correlation
$$
C_T(\tau)  \stackrel{T \to \infty}{\longrightarrow} C^{(1)}(\tau)  ,
$$
in probability, where
\begin{eqnarray}
\nonumber
C^{(1)}(\tau)= \frac{1}{(2\pi)^2} \sum_{k =-\infty}^\infty
 \int_{-\infty}^\infty \iint_{[0,2\pi]^2}
 \overline{\hat{G}(\omega,{\bx_1},{\br}_\theta)} \hat{G}( \omega,{\bx_2},{\br}_{\theta'}) 
 \frac{1}{v} F \Big( \frac{\theta'-\theta-2\pi k}{v} \Big) \\
 \times
\exp\Big( i \frac{\omega}{v}(\theta'- \theta-2\pi k) \Big)d\theta d\theta'
e^{-i \omega \tau}   d \omega ,
\label{eq:C1aprop1}
\end{eqnarray}
${\br}_\theta=(R_0\cos \theta ,R_0\sin \theta )$ and $\hat{G}$ is the Green's function solution to
(\ref{eq:greenhetero})
with Sommerfeld radiation condition.
If, additionally, Assumption (H2) is satisfied:
$$
\mbox{\rm (H2)}
 \hspace*{0.8in}
 \begin{array}{l}
 \mbox{the time for a loop $2\pi /v$  is much larger}\\
 \mbox{than the coherence time of the source},
 \end{array}
 \hspace*{1.45in}
$$
then 
\begin{eqnarray}
\nonumber
C^{(1)}(\tau)= \frac{1}{(2\pi)^2}
 \int_{-\infty}^\infty \iint_{[0,2\pi]^2}
 \overline{\hat{G}(\omega,{\bx_1},{\br}_\theta)} \hat{G}( \omega,{\bx_2},{\br}_{\theta'}) 
 \frac{1}{v} F \Big( \frac{\theta'-\theta}{v} \Big) \\
 \times
\exp\Big( i \frac{\omega}{v}(\theta'- \theta) \Big)d\theta d\theta'
e^{-i \omega \tau}   d \omega .
\label{eq:C1bprop1}
\end{eqnarray}
\end{proposition}
Note that the number of loops carried out by the source is (the integer part of) $2\pi v T$.
So the condition $vT \gg 1$ for statistical stability means that the source has to make many loops.

{\it Proof.}
In the Fourier domain, the source has the form
$$
\hat{s}(\omega,{\bx}) = 
\frac{1}{R_0} \delta(r-R_0) \hat{n}(\omega,\theta)  ,
$$
where we use the polar coordinates ${\bx}=(r\cos \theta ,r\sin \theta)$,
and
$$
\hat{n}(\omega,\theta) = \frac{1}{v} \sum_{k=-\infty}^\infty \exp\Big(i \frac{\omega}{v} (\theta-2k\pi)\Big)
f \Big( \frac{\theta-2\pi k}{v}\Big) .
$$
The second-order moment is
\begin{eqnarray*}
\EE \big[ \overline{\hat{n}(\omega,\theta)} \hat{n}(\omega',\theta') \big] = \frac{1}{v^2}
 \sum_{k,k'=-\infty}^\infty \exp\Big( i \frac{\omega'}{v} (\theta'- 2k' \pi) - i \frac{\omega}{v} (\theta-2k\pi) \Big)\\
 \times
F \Big( \frac{\theta'-\theta-2\pi (k'-k)}{v}\Big) .
\end{eqnarray*}
Using the Poisson summation formula $\sum_k e^{ i 2\pi k s} = \sum_k \delta(s-k)$ this can also be written as:
\begin{eqnarray*}
\EE \big[ \overline{\hat{n}(\omega,\theta)} \hat{n}(\omega',\theta') \big] = \frac{1}{2\pi v}
 \sum_{k,k'=-\infty}^\infty \delta \big(\omega-\omega'-kv\big) 
  \exp\Big( i \frac{\omega' (\theta ' -2\pi k')-\omega\theta}{v}   \Big) \\
  \times
F \Big( \frac{\theta'-\theta-2\pi k'}{v}\Big) .
\end{eqnarray*}
The cross correlation (\ref{def:CT0})
 can be expressed in terms of the Fourier components of the recorded signals as
\begin{eqnarray*}
C_T(\tau) = \frac{1}{(2\pi)^2}\int_{-\infty}^\infty \int_{-\infty}^\infty \overline{\hat{u}(\omega,{\bx_1})} \hat{u}( \omega',{\bx_2}) 
{\rm sinc} \Big[ \frac{ T(\omega-\omega')}{2} \Big] \\
\times
\exp\Big( -i  \frac{ T(\omega-\omega')}{2}\Big)  d\omega e^{-i \omega' \tau} d \omega'  .
\end{eqnarray*}
In terms of the Green's function the wave field is (\ref{eq:expressuG}), which gives
$$
\hat{u}(\omega,{\bx}) =
\int_0^{2\pi} \hat{G}(\omega,{\bx},{\br}_\theta) \hat{n}(\omega,\theta) d \theta ,
$$
with ${\br}_\theta=(R_0\cos \theta ,R_0\sin \theta )$.
Taking the expectation and using $vT \gg 1$,  we find
$$
\EE \big[ C_T(\tau) \big] = C^{(1)}(\tau) ,
$$
with $C^{(1)}$ given by (\ref{eq:C1aprop1}).
One can also compute the variance of $C_T(\tau)$ and show that it is of order $1/T$ as in \cite{garnier09},
which gives the first result of the proposition.\\
If, additionally, $2\pi /v$  is much larger than the coherence time of the source, then only the term $k=0$ contributes 
to leading order and the second result is proved.
\hfill {\small $\Box$}\\

{\bf Analysis for a white noise source model.}
In the regime in which the noise sources are delta-correlated in time (the white-noise approximation) $F(t)=\delta(t)$,
then, by (\ref{eq:C1bprop1}), the Fourier transform of the statistical cross correlation 
$$
\hat{C}^{(1)}(\omega)= \int_{-\infty}^{\infty} C^{(1)}(\tau) e^{i \omega \tau} d\tau 
$$
is related to the Green's function through the relation:
$$
\hat{C}^{(1)}(\omega)= \frac{1}{2\pi} \int_{0}^{2\pi}
 \overline{\hat{G}(\omega,{\bx_1},{\br}_\theta)} \hat{G}( \omega,{\bx_2},{\br}_{\theta})  d\theta   .
$$
By Helmholtz-Kirchhoff identity (see, for instance \cite[p. 419]{born} or \cite[Theorem 2.33]{ammari}) we have
$$
 \int_{0}^{2\pi}
 \overline{\hat{G}(\omega,{\bx_1},{\br}_\theta)} \hat{G}( \omega,{\bx_2},{\br}_{\theta}) 
R_0 d\theta  = \frac{c_0}{\omega} {\rm Im} \big\{  \hat{G}( \omega,{\bx_1},{\bx_2}) \big\} ,
$$
therefore we find 
\begin{equation}
\label{eq:classicc20}
\hat{C}^{(1)}(\omega)= \frac{1}{2\pi R_0} \frac{c_0}{\omega}{\rm Im} \big\{  \hat{G}( \omega,{\bx_1},{\bx_2}) \big\}  .
\end{equation}
This formula is classical by comparison with the situation in which there are fixed point sources 
at the perimeter of the disk with radius $R_0$ (with unit density)  that emit uncorrelated stationary signals \cite{garnier09,wap10}. 
In this situation  the cross correlation of the recorded signals is 
$$
\hat{C}^{(1)}(\omega)= \frac{c_0}{\omega}  {\rm Im}\big\{ \hat{G}(\omega,{\bx_1},{\bx_2}) \big\}  .
$$
The factor $1/(2\pi R_0)$ in (\ref{eq:classicc20}) can be interpreted from the fact that the average density of sources
in the case of a point source moving along the circular trajectory with radius $R_0$ is precisely $1/(2\pi R_0)$.\\

{\bf Analysis for a homogeneous medium and for $R_0 \gg |{\bx_1}|,|{\bx_2}|$.}
Here we do not assume that the noise source is a white noise, but we assume that the medium is homogeneous $c(\bx)\equiv c_0$, 
so that the Green's function is equal to (\ref{def:green0}).
The following proposition gives the exact relationship between the statistical cross correlation of the noise signals recorded 
at the two observation points and the imaginary part of the Green's function between them.

\begin{proposition}
\label{prop:0}%
Under the assumption (H2), if the medium is homogeneous with background velocity $c_0$ and if  $R_0 \gg |{\bx_1}|,|{\bx_2}|$, then the Fourier transform of the statistical cross correlation is given by:
\begin{equation}
\hat{C}^{(1)}(\omega)=\frac{1}{ 2\pi R_0} \frac{c_0}{\omega}  
\int_{-\infty}^\infty e^{-i \omega t} F(t) 
{\rm Im} \Big\{ \hat{G}_0\Big(\omega \sqrt{1+\frac{v^2 t^2}{4}},{\bx_1},{\bx_2}\Big)  \Big\} dt  .
\label{eq:C1sec1}
\end{equation}
\end{proposition}

{\it Proof.} 
After the change of variables $(\theta,\theta') \to (\theta+h/2,\theta-h/2)$ we have
\begin{eqnarray*}
\hat{C}^{(1)}(\omega)= \frac{1}{2\pi}
\int_0^{2\pi} \int_{\max(-2\theta , 2\theta-4\pi)}^{\min(2\theta,4\pi-2\theta)}
 \overline{\hat{G}_0(\omega,{\bx_1},{\br}_{\theta+h/2})} \hat{G}_0( \omega,{\bx_2},{\br}_{\theta-h/2})  \\
 \times
 \frac{1}{v} F \Big( \frac{h}{v} \Big)
\exp\Big( -  i \frac{\omega}{v}h \Big) dh d\theta .
\end{eqnarray*}
In the regime $R_0 \gg |{\bx_1}|,|{\bx_2}|$ we can write
$$
\hat{G}_0(\omega,{\bx_j},{\br}_{\theta+h/2}) \simeq \hat{G}_0(\omega,{\bx_j},{\br}_\theta)
\exp \Big( i \frac{\omega}{c_0} \big( |{\br}_{\theta+h/2} -{\bx_j}| - |{\br}_{\theta}-{\bx_j}| \big) \Big) ,
$$
for $j=1,2$,
and therefore $\hat{C}^{(1)}(\omega)$ has the form
\begin{eqnarray*}
\hat{C}^{(1)}(\omega)&=&  \frac{1}{2\pi} \int_0^{2\pi} \int_{\max(-2\theta , 2\theta-4\pi)}^{\min(2\theta,4\pi-2\theta)} \overline{\hat{G}_0(\omega,{\bx_1},{\br}_\theta)} \hat{G}_0( \omega,{\bx_2},{\br}_{\theta}) 
 \frac{1}{v} F \Big( \frac{h}{v} \Big)\\
 &&\times
\exp\Big( - i \frac{\omega}{v}h -
 i \frac{\omega}{c_0} \big( |{\br}_{\theta+h/2} -{\bx_1}| - |{\br}_{\theta}-{\bx_1}| \big)
 \\
 &&\hspace*{0.5in} +
 i \frac{\omega}{c_0} \big( |{\br}_{\theta-h/2} -{\bx_2}| - |{\br}_{\theta}-{\bx_2}|\big) \Big)
dh d\theta  .
\end{eqnarray*}
Using the assumption (H2) that $2\pi /v$ is much larger than the coherence time of the source,
this can be simplified as
\begin{eqnarray*}
\hat{C}^{(1)}(\omega)&=&  \frac{1}{2\pi} \int_0^{2\pi} \int_{-\infty}^\infty
\overline{\hat{G}_0(\omega,{\bx_1},{\br}_\theta)} \hat{G}_0( \omega,{\bx_2},{\br}_{\theta}) 
F (t )\\
 &&\times
\exp\Big( - i  \omega t -
 i \frac{\omega}{c_0} ( |{\br}_{\theta+vt/2} -{\bx_1}| - |{\br}_{\theta}-{\bx_1}|)
\\
 &&\hspace*{0.5in} +
 i \frac{\omega}{c_0} ( |{\br}_{\theta-vt/2} -{\bx_2}| - |{\br}_{\theta}-{\bx_2}|) \Big)
dt d\theta  .
\end{eqnarray*}
Furthermore, writing $\bx_j=(x_j,y_j)$,
$$
 |{\br}_{\theta+vt/2} -{\bx_j}| - |{\br}_{\theta}-{\bx_j}| \simeq 
 \frac{vt}{2} \frac{\partial}{\partial \theta} |{\br}_{\theta}-{\bx_j}| \simeq \frac{vt}{2}
 ( x_j \sin \theta- y_j \cos \theta)  ,
$$
since $R_0 \gg |{\bx_j}|$, $j=1,2$.
As a result:
\begin{eqnarray*}
\hat{C}^{(1)}(\omega)&=&
 \frac{1}{2\pi} \int_{0}^{2\pi} \int_{-\infty}^{\infty} 
 \overline{\hat{G}_0(\omega,{\bx_1},{\br}_\theta)} \hat{G}_0( \omega,{\bx_2},{\br}_{\theta}) 
 F(t)  \\
 &&
 \times
 \exp\Big( - i \omega t + i \omega  \frac{vt}{2c_0} \big( (x_2-x_1) \sin \theta-(y_2-y_1) \cos \theta \big) \Big)
dt 
d\theta .
\end{eqnarray*}
Using the asymptotic form of the Hankel function
\begin{equation}
\label{eq:asymptHankel}
H_0^{(1)}(s) \stackrel{s \gg 1}{\simeq} \frac{\sqrt{2}}{\sqrt{\pi s}} e^{i s -i \frac{\pi}{4}},
\end{equation}
and the expansion (valid when $R_0 \gg |{\bx_1}|,|{\bx_2}|$)
$$
|{\bx_2}-{\br}_\theta| - |{\bx_1}-{\br}_\theta| \simeq -(x_2-x_1) \cos \theta  -(y_2-y_1) \sin \theta  ,
$$
we finally get the desired result from the identity $\int_0^{2\pi } e^{ i s \sin\theta} d\theta=2\pi J_0(s)$.
\hfill {\small $\Box$}\\

Let us discuss the results of Proposition \ref{prop:0}.
In the regime in which the noise sources are delta-correlated in time (the white-noise approximation) $F(t)=\delta(t)$,
or the noise sources have positive finite coherence time, but
the time $2\pi/v$ for a loop is much larger  than the coherence time of the noise source and than 
the travel time $|{\bx_1}-{\bx_2}|/c_0$ from ${\bx_1}$ to ${\bx_2}$, then we recover the classical form:
\begin{equation}
\label{eq:classicc2}
\hat{C}^{(1)}(\omega)=\frac{1}{ 2\pi R_0} \frac{c_0}{\omega}  
\hat{F}(\omega ) {\rm Im}\big\{ \hat{G}_0(\omega,{\bx_1},{\bx_2}) \big\}  .
\end{equation}
In the general case, Eq.~(\ref{eq:C1sec1}) shows that the relation between the statistical cross correlation and the
imaginary part of the Green's function is affected by a Doppler-like effect (i.e. a frequency shift).
It is interesting to find the correction to the classical formula when the velocity of the source is large
enough so that 
$$
M_{1,2} = \frac{v_0}{c_0} \frac{|{\bx_1}-{\bx_2}|}{2R_0}  
$$
is smaller than one but not vanishing. Note that,
since $R_0 \gg |{\bx_1}|,|{\bx_2}|$, this requires that $v_0 \gg c_0$, which seems a quite extreme supersonic regime. 
But we will see that, even in these extreme conditions, travel time estimation can be successfully carried out.
After some algebra we find from (\ref{eq:C1sec1})
\begin{eqnarray}
\nonumber
\frac{\partial C^{(1)}}{\partial \tau} (\tau) = 
-\frac{c_0}{ 8 \pi^2 R_0 \sqrt{ 1-M_{1,2} ^2}}
\int_0^\infty \hat{F}(\omega) 
\sin \Big( \frac{\omega}{1- M_{1,2}^2} \tau \Big) \\
\times
J_0 \Big( \omega 
\sqrt{\frac{|{\bx_1}-{\bx_2}|^2}{ c_0^2(1- M_{1,2}^2)} + \frac{M_{1,2}^2 \tau^2}{ (1-M_{1,2}^2)^2}}
\Big) d\omega .
\end{eqnarray}
This expression depends on the velocity $v_0$ via the term $M_{1,2}$, 
which is a manifestation of the Doppler effect.
However, travel time estimation based on this Green's function estimation
is not affected by the velocity of the source as shown by the following arguments.

In the context of travel time estimation, we consider a situation in which the travel time
$|{\bx_1}-{\bx_2}|/c_0$ is  larger than the coherence time of the source. 
When the motion of the source has small velocity, so that $M_{1,2} \ll 1$,
the cross correlation (as well as the Green's function) has a peak at time lag $\tau$ 
equal to the travel time $|{\bx_1}-{\bx_2}|/c_0$, and the width of the peak is conversely
proportional to the noise bandwidth.
This is still true in the case of a moving source.
Indeed, under assumption (H1),
we have
\begin{eqnarray*}
\frac{\partial C^{(1)}}{\partial \tau} (\tau) &\simeq &
-\frac{c_0^{3/2} }{ 2  (2\pi)^{3/2} R_0 \sqrt{ \omega_0 |{\bx_1}-{\bx_2}|}} \\
&&\times
F_{\rm B} \Big(  \frac{\tau}{1- M_{1,2}^2} 
- \sqrt{\frac{|{\bx_1}-{\bx_2}|^2}{ c_0^2(1- M_{1,2}^2)} + \frac{M_{1,2}^2 \tau^2}{ (1-M_{1,2}^2)^2}}
\Big) \\
&&
\times \sin \Big( \omega_0 \big( \frac{\tau}{1- M_{1,2}^2} 
- \sqrt{\frac{|{\bx_1}-{\bx_2}|^2}{ c_0^2(1- M_{1,2}^2)} + \frac{M_{1,2}^2 \tau^2}{ (1-M_{1,2}^2)^2}} \big) +\frac{\pi}{4}
 \Big),
\end{eqnarray*}
for $\tau >0$. This shows that the cross correlation has a peak at time lag $\tau_{\rm max}$
such that the argument inside $F_{\rm B}$ is zero, that is, 
$$
\frac{\tau_{\rm max}}{1- M_{1,2}^2} 
= \sqrt{\frac{|{\bx_1}-{\bx_2}|^2}{ c_0^2(1- M_{1,2}^2)} + \frac{M_{1,2}^2 \tau_{\rm max}^2}{ (1-M_{1,2}^2)^2}} ,
$$
which is exactly 
$$
\tau_{\rm max} = \frac{|{\bx_1}-{\bx_2}|}{c_0} ,
$$
 independently of $v_0$.
Therefore the peak is at time lag equal to the travel time from $\bx_1$ to $\bx_2$.
Around time lag $\tau_{\rm max}$, the cross correlation has the form:
\begin{eqnarray*}
\frac{\partial C^{(1)}}{\partial \tau} (\tau_{\rm max} + \tau) &\simeq &
-\frac{c_0^{3/2} }{ 2  (2\pi)^{3/2} R_0 \sqrt{ \omega_0 |{\bx_1}-{\bx_2}|}} \\
&&\times
F_{\rm B} \Big( \tau \big(1-\frac{M_{1,2} v_0}{4R_0}\tau\big) \Big) \sin \Big( \omega_0 
  \tau \big(1-\frac{M_{1,2} v_0}{4R_0}\tau\big) +\frac{\pi}{4}
 \Big).
\end{eqnarray*}
This  shows that:
\begin{enumerate}
\item
 the carrier frequency of the cross correlation around time lag $\tau_{\rm max}$ is $\omega_0$, 
\item
 travel time estimation with the empirical cross correlation with a moving random source is unbiased
(i.e., with  a bias smaller than the resolution),
\item
 travel time estimation has the same resolution as in the case with a set of stationary noise 
sources surrounding the two receivers at ${\bx_1}$ and ${\bx_2}$.
\end{enumerate}

\section{Conclusions}
In this paper we have investigated the possibility to use moving sensors to create large synthetic apertures 
in ambient noise correlation-based imaging. 
We consider in this paper situations with periodic trajectories 
so that it is possible to achieve statistical stability for the cross correlation of the recorded signals
(i.e. the empirical cross correlation is approximately equal to the statistical cross correlation).
We were motivated by 1) the recent result that time-reversal refocusing for a moving source is possible and 
resolution enhancement is observed when the source velocity becomes non-negligible compared to the wave speed, and 
2) the classical analogy between time reversal and correlation-based imaging. However this analogy
is broken when the sensors are moving, because the lack of source-receiver reciprocity and the Doppler effects 
do not play the same role in the two situations. As a consequence, it is possible to carry out correlation-based imaging provided the 
sensor velocity is small compared to the wave speed. When the sensor  velocity becomes non-negligible compared to the wave speed,
then it is necessary to build carefully designed imaging functions to avoid localization bias due to Doppler effect
but resolution is then reduced
compared to the case of small velocity. These modified imaging functions depend on the trajectory of the moving receiver.

We have presented a few ideas to perform synthetic experiments to check the theoretical predictions at the ends of the sections.
Real experiments could be carried out in the framework of water waves, for which interesting time-reversal experiments
have recently been carried out, and that would allow to consider motions with large speeds (large relative to the speed of propagation) \cite{bacot}.

\appendix

\section{Comparison with time reversal}
\label{app:TR}%
 
The efficiency of correlation-based imaging is classically explained by its time-reversal interpretation \cite{derode,garnier16}.
However, when the sources or receivers are moving, the analogy is not so clear.
The goal of this section is to study the time-reversal experiment that should be the analogous 
of the correlation experiment described in Section \ref{sec:gre} and to show that the results 
are different.\\

We consider a point source moving on a circular trajectory. Its position is 
${\bx}_{\rm s}(t) = (R_0 \cos(vt), R_0 \sin(vt))$, where $R_0>0$ is the radius
of its circular trajectory and $v$ is its angular velocity (its linear velocity is $v_0=vR_0$).
It emits the pulse $f(t)$, whose support is in the time interval $[-\pi /v, \pi/v]$
(which means the emission occurs during a single loop).
A time-reversal mirror is located at the surface of a large ball $B$ 
(the ball $B$ does not need to be centered at ${\bf 0}$, but it needs to enclose the circular trajectory of the source).

\begin{figure}
\begin{center}
\begin{tabular}{c}
 \psfig{file=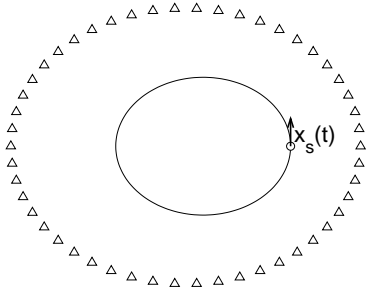,width=6.0cm}
\end{tabular}
\end{center}
\caption{Experimental set-up for the time-reversal experiment in Appendix \ref{app:TR}. 
The source ${\bx}_{\rm s}(t)$ is moving 
on a circular trajectory (with radius $R_0$) and the triangles are the sources/receivers of the time-reversal mirror
(on $\partial B$).
\label{fig2}}
\end{figure}

The source term is
$$
s(t,{\bx}) = \delta( {\bx} - {\bx}_{\rm s}(t) )  f(t) .
$$
In the Fourier domain and using polar coordinates ${\bx}=(r\cos \theta,r\sin \theta)$, we have
$$
\hat{s}(\omega,{\bx}) = \frac{1}{v_0} \delta(r-R_0) f\Big( \frac{\theta}{v} \Big) \exp \Big( i \omega \frac{\theta}{v} \Big) .
$$
The signal recorded by the time-reversal mirror at ${\bx} \in \partial B$ is 
\begin{eqnarray*}
\hat{u}(\omega,{\bx}) &=& \int_{\RR^2} \hat{G}_0(\omega,{\bx},{\bx'}) \hat{s}(\omega,{\bx'}) d{\bx'} \\
&=&
\frac{1}{v_0}
\int_0^{2\pi} f\Big( \frac{\theta}{v} \Big) \exp \Big( i \omega \frac{\theta}{v} \Big) \hat{G}_0( \omega,{\bx},{\br}_\theta) d\theta ,
\end{eqnarray*}
where ${\br}_\theta=(R_0 \cos \theta, R_0 \sin\theta)$ and $\hat{G}_0$ is the homogeneous Green's function (\ref{def:green0}).
After time-reversal, the signal that refocuses at a point ${\br}_{\theta}$, $\theta \in [0,2\pi)$, on the circular trajectory of the original source is
\begin{eqnarray}
\nonumber
\hat{u}_{\rm TR}( \omega , {\br}_{\theta} ) &=& \int_{\RR^2} \frac{c_0}{\omega} 
{\rm Im}\big\{ \hat{G}_0(\omega, {\br}_{\theta},{\bx'}) \big\}
\hat{s}(\omega,{\bx'}) d{\bx'} \\
&=&
\frac{c_0}{v_0}
\int_0^{2\pi} \frac{1}{\omega}
f\Big( \frac{\theta'}{v} \Big) \exp \Big( i \omega \frac{\theta'}{v} \Big){\rm Im} \big\{ \hat{G}_0( \omega,{\br}_{\theta},{\br}_{\theta'}) \big\} d\theta .
\label{eq:TRprofile1}
\end{eqnarray}

Let us compare the time-reversed refocused wave (\ref{eq:TRprofile1}) with the cross-correlation formula (\ref{eq:corrstata1}).
For simplicity we assume (H1) and the equivalent assumption for the time-reversal experiment:
the source is of the form $f(t)=e^{-i \omega_0 t} f_{\rm B}(t) +c.c.$.
Then, using the result in \cite[Sec. 4]{garfink},
we find that
\begin{eqnarray*}
{u}_{\rm TR}( t, {\br}_{\theta} ) =- \frac{c_0}{4 \sqrt{1-M^2} \omega_0} 
\exp \Big( - i\omega_0 \frac{t + M \frac{R_0}{c_0} \theta}{1-M^2} \Big) f_{\rm B}\Big( -\frac{t + M \frac{R_0}{c_0} \theta}{1-M^2}\Big) \\
\times
J_0 \Big( \frac{\omega_0 R_0 |\theta+vt|}{c_0(1-M^2)} \Big) +c.c.,
\end{eqnarray*}
with $M=v_0/c_0\in (0,1)$. Note that the factor $1-M^2$ in the Bessel function $J_0$ shows that resolution is enhanced when the source 
velocity becomes comparable to the wave speed.
Thus we can see that both expressions involve the imaginary part of the Green's function, but the corrections due to the 
velocity $v$ are different.

\section{Corrective term to the imaging function}
\label{app:B}%
We complete the analysis of the imaging functions carried out in Section \ref{sec:ref}.
When $R_0 \gg |{\by}_{\rm ref}|$, we have for $k \geq 1$: 
\begin{eqnarray*}
\frac{\partial^{2k}}{\partial \theta^{2k}}  |{\br}_\theta-{\by}_{\rm ref}| &=& -  (-1)^{k} \big(x_{\rm ref} \cos \theta
+ y_{\rm ref} \sin \theta \big)  \\
&&- \frac{1}{4} (-4)^k
 \frac{ (x_{\rm ref}^2 - y_{\rm ref}^2)\cos 2\theta 
 + 2 x_{\rm ref} y_{\rm ref} \sin 2\theta}{R_0}
 + O \Big( \frac{|{\by}_{\rm ref}|^3}{R_0^2} \Big)  ,
\end{eqnarray*}
and therefore we get
\begin{eqnarray*}
&&
|{\br}_{\theta+h/2}-{\by}_{\rm ref}| + |{\by}_{\rm ref}-{\br}_{\theta-h/2}| \\
&&= 
2 |{\br}_\theta-{\by}_{\rm ref}|  -
2 \sum_{k=1}^\infty \frac{(-1)^k}{(2k) !} \Big( \frac{h}{2} \Big)^{2k}  \big(x_{\rm ref} \cos \theta
+ y_{\rm ref} \sin \theta \big)  \\
&&\quad
-
\frac{1}{2} \sum_{k=1}^\infty \frac{(-1)^k}{(2k) !} h^{2k}\frac{ (x_{\rm ref}^2 - y_{\rm ref}^2)\cos 2\theta 
 + 2 x_{\rm ref} y_{\rm ref} \sin 2\theta}{R_0}
 + O \Big( \frac{|{\by}_{\rm ref}|^3}{R_0^2} \Big) 
\\
&&=2 |{\br}_\theta-{\by}_{\rm ref}|  +
4  \sin^2 \Big( \frac{h}{4} \Big)   \big(x_{\rm ref} \cos \theta
+ y_{\rm ref} \sin \theta \big)  \\
&&\quad  + \sin^2 \Big( \frac{h}{2} \Big)   \frac{ (x_{\rm ref}^2 - y_{\rm ref}^2)\cos 2\theta 
 + 2 x_{\rm ref} y_{\rm ref} \sin 2\theta}{R_0}
 + O \Big( \frac{|{\by}_{\rm ref}|^3}{R_0^2} \Big) 
,
\end{eqnarray*}
and finally:
\begin{eqnarray}
\nonumber
C^{(1)}_1 \Big( \theta+\frac{h}{2} ,\theta - \frac{h}{2} \Big) 
&=& 
\frac{\sigma_{\rm ref}}{16\pi R_0} F \Big( \frac{h}{v} -2 \frac{|{\br}_\theta -{\by}_{\rm ref}|}{c_0} -  \frac{D(h,\theta)}{c_0}
\Big)  \\
&&
  + \frac{\sigma_{\rm ref}}{16\pi R_0} F \Big( \frac{h}{v} 
  +2 \frac{|{\br}_\theta -{\by}_{\rm ref}|}{c_0} + \frac{D(h,\theta)}{c_0}
\Big) ,
\end{eqnarray}
with
\begin{eqnarray}
\nonumber
D(h,\theta)&=& 4 \sin^2 \big( \frac{h}{4}\big)
\big(x_{\rm ref}\cos \theta  + y_{\rm ref} \sin \theta\big) \\
&&
+   \sin^2\big(\frac{h}{2} \big) \frac{ (x_{\rm ref}^2 - y_{\rm ref}^2)\cos 2\theta 
 + 2 x_{\rm ref} y_{\rm ref} \sin 2\theta}{R_0}  +  O \Big( \frac{|{\by}_{\rm ref}|^3}{R_0^2} \Big)  .
\end{eqnarray}
We have
\begin{eqnarray*}
&&D\Big( 2\frac{v}{c_0} |{\br}_\theta-{\by}^S|, \theta\Big) \\
&&= 
4 \sin^2 \Big(\frac{v_0}{2c_0}\Big) \big(x_{\rm ref}\cos \theta  + y_{\rm ref} \sin \theta \big) \\
&&\quad 
- \sin \Big(\frac{v_0}{c_0}\Big) \frac{2v_0}{c_0}  \frac{ (x_{\rm ref}\cos \theta  + y_{\rm ref} \sin \theta) (x^S \cos \theta  + y^S \sin \theta) }{ R_0}  \\
&&\quad 
+  \sin^2 \Big(\frac{v_0}{c_0}\Big)\frac{ (x_{\rm ref}^2 - y_{\rm ref}^2)\cos 2\theta 
 + 2 x_{\rm ref} y_{\rm ref} \sin 2\theta}{ R_0} +  O \Big( \frac{|{\by}_{\rm ref}|^3}{R_0^2} \Big) ,
 \end{eqnarray*}
and therefore, for ${\by}^S$ in a neighborhood of ${\by}_{\rm ref} \cos (v_0/c_0)$,
\begin{eqnarray*}
&& 2\frac{ |{\br}_\theta-{\by}^S|}{c_0} - 2\frac{ |{\br}_\theta-{\by}_{\rm ref}|}{c_0} - D\Big( 2\frac{v}{c_0} |{\br}_\theta-{\by}^S|, \theta\Big)\\
&&= 
2 \Big( \cos\big( \frac{v_0}{c_0}\big) x_{\rm ref} - x^S\Big) \cos \theta 
+
2 \Big( \cos\big( \frac{v_0}{c_0}\big) y_{\rm ref} - y^S\Big) \sin \theta \\
&&\quad 
+
\frac{(x_{\rm ref} \cos \theta -y_{\rm ref} \sin \theta)^2}{R_0} \Big( \frac{v_0}{c_0} \sin \Big( \frac{2v_0}{c_0} \big) - \sin^2 \big( \frac{v_0}{c_0} \big)\Big)
+  O \Big( \frac{|{\by}_{\rm ref}|^3}{R_0^2} \Big) .
 \end{eqnarray*}

This shows that the main peak in the imaging function tends to disappear when 
$({v_0^2}/{c_0^2})[ {|{\by}_{\rm ref}|^2}/({R_0 \lambda})]$ becomes of order one.
In other words, the imaging function is valid to image reflectors within the disk of center ${\bf 0}$ (the 
center of the circular trajectory of the receiver) and radius $\sqrt{R_0 \lambda} (c_0 / v_0)$.


\begin{thebibliography}{9}

\bibitem{ammari}
H. Ammari, J. Garnier, W. Jing, H. Kang, M. Lim, K. S\o lna, and H. Wang,
{\it Mathematical and Statistical Methods for Multistatic Imaging},
Lecture Notes in Mathematics, Vol. 2098, Springer, Berlin, 2013.

\bibitem{bacot}
V. Bacot, M. Labousse, A. Eddi, M. Fink, and E. Fort,
Time reversal and holography with spacetime transformations,
Nature Physics {\bf 12} (2016), pp.~972--977.

\bibitem{badon15}
A. Badon, G. Lerosey, A. C. Boccara, M. Fink, and A. Aubry, 
Retrieving time-dependent Green's functions in optics with low-coherence interferometry, 
Phys. Rev. Lett. {\bf 114} (2015),  023901.
  
\bibitem{bardos08}
{C. Bardos, J. Garnier, and G. Papanicolaou},
{Identification of  Green's functions singularities by cross correlation of noisy signals},
{Inverse Problems} {\bf 24} (2008), 015011.

\bibitem{born}
M. Born and E. Wolf,
{\it Principles of Optics},
Cambridge University Press, Cambridge, 1999.

\bibitem{campillo03}
M. Campillo and A. Paul, 
Long-range correlations in the diffuse seismic coda, 
Science {\bf 299} (2003), pp.~547--549.

\bibitem{davy}
M. Davy, M. Fink, and J. de Rosny, 
Green's function retrieval and passive imaging from correlations of wideband thermal radiations, 
Phys. Rev. Lett. {\bf 110} (2013), 203901.

\bibitem{derode}
A. Derode, E. Larose, M. Campillo, and M. Fink, 
How to estimate the Green's function of a heterogeneous medium between two passive sensors~? Application
to acoustic waves,
Appl. Phys. Lett. {\bf 83} (2003), 3054--3056.


\bibitem{brenguier08}
{F. Brenguier, N. M. Shapiro, M. Campillo, V.  Ferrazzini, Z. Duputel, O. Coutant, and A. Nercessian}, 
{Towards forecasting volcanic eruptions using seismic noise},
{Nature Geoscience} {\bf 1} (2008), 126--130.

\bibitem{colin09}
{Y. Colin de Verdi\`ere},
{Semiclassical analysis and passive imaging},
{Nonlinearity} {\bf 22} (2009), R45--R75.

\bibitem{garnier05}
J. Garnier,
Imaging in randomly layered media by cross-correlating noisy signals,
SIAM Multiscale Model. Simul. {\bf 4} (2005), 610--640.

\bibitem{garfink}
J. Garnier and M. Fink, 
Super-resolution in time-reversal focusing on a moving source, 
Wave Motion {\bf 53} (2015),  80--93.

\bibitem{garnier09}
J. Garnier and G. Papanicolaou, 
Passive sensor imaging using cross correlations of noisy signals in a scattering medium,
SIAM J. Imaging Sciences {\bf 2} (2009), 396--437. 

\bibitem{garnier10}
J. Garnier and G. Papanicolaou, 
Resolution analysis for imaging with noise, 
Inverse Problems {\bf 26} (2010), 074001.

\bibitem{garnier16}
J. Garnier and G. Papanicolaou, 
{\it Passive Imaging with Ambient Noise}, 
Cambridge University Press, Cambridge, 2016.

\bibitem{gouedard08}
{P. Gou\'edard, L. Stehly, F. Brenguier, M. Campillo, Y. Colin de Verdi\`ere, E. Larose,  L. Margerin, P. Roux, F. J. Sanchez-Sesma, N. M. Shapiro, and R. L. Weaver},
{Cross-correlation of random fields: mathematical approach and applications},
{Geophysical Prospecting} {\bf 56} (2008), 375--393.

\bibitem {roux05}
{P. Roux, K. G. Sabra, W. A. Kuperman, and A. Roux},
{Ambient noise cross correlation in free space: Theoretical approach},
{J. Acoust.  Soc. Am.} {\bf 117} (2005), 79--84.


\bibitem{sabra10}
K. G. Sabra,
Influence of the noise sources motion on the estimated Green's functions from ambient noise cross-correlations,
J. Acoust. Soc. Am. {\bf 127} (2010), 3577--3589.


\bibitem{schuster}
G. T. Schuster,
{\it Seismic Interferometry},
Cambridge University Press, Cambridge, 2009.

\bibitem{shapiro05}
{N. M. Shapiro, M. Campillo, L. Stehly,  and M. H. Ritzwoller},
{High-resolution surface wave tomography from ambient noise},
{Science} {\bf 307} (2005), 1615--1618.


\bibitem{snieder04}
{R. Snieder},
{Extracting the Green's function from the correlation of coda waves:
A derivation based on stationary phase},
{Phys. Rev. E} {\bf 69} (2004), 046610.

\bibitem{wap04}
{K. Wapenaar},
{Retrieving the elastodynamic Green's function of an arbitrary 
inhomogeneous medium by cross correlation},
Phys. Rev. Lett. {\bf 93} (2004), 254301.

\bibitem{wap10}
K. Wapenaar, E. Slob, R. Snieder, and A. Curtis,
Tutorial on seismic interferometry:
Part 2 - Underlying theory and new advances,
Geophysics {\bf 75} (2010), 75A211--75A227.

\bibitem{weaver}
{R. Weaver and O. I. Lobkis},
{Ultrasonics without a source: Thermal fluctuation
correlations at MHz frequencies},
{Phys. Rev. Lett.} {\bf 87} (2001), 134301.


\bibitem{dehoop06}
{H. Yao, R. D. van der Hilst, and M. V. de Hoop},
{Surface-wave array tomography in SE Tibet from ambient seismic noise and two-station analysis I. Phase velocity maps},
{Geophysical Journal International} {\bf 166} (2006), 732--744.

\bibitem{pied}
{Throughout the paper,
symbols of scalar quantities are printed in italic type and
symbols of vectors are printed in bold italic type.}

\end{thebibliography}
\end{document}